\documentclass[prb,twocolumn,showpacs]{revtex4}
\usepackage{color}
\usepackage{bbm}
\usepackage{units}
\usepackage{graphics}
\usepackage{natbib}
\input boxedeps.tex
\input boxedeps.cfg
\HideDisplacementBoxes

\begin{document}
\title{Lifshitz transitions and elastic properties of Osmium under pressure}
\author{Daniela Koudela, Manuel Richter, Arnulf M\"obius, Klaus Koepernik,
and Helmut Eschrig}
\affiliation{IFW Dresden, PF 270 116, D-01171 Dresden, Germany}
\begin{abstract}
Topological changes of the Fermi surface under pressure may
cause anomalies in the low-temperature
elastic properties.
Our density functional calculations for elemental Osmium evidence that
this metal undergoes three such
Lifshitz transitions in the pressure range between 70 GPa and 130 GPa.
The related elastic anomalies are, however, invisibly weak.
The critical pressures considerably exceed the values for
recently measured and calculated anomalies in the
pressure $(P)$ dependence of the hexagonal $c/a$ lattice parameter ratio close
to 25 GPa. We demonstrate that the latter anomalies are statistically
not significant and that $(c/a)(P)$ can be fitted equally well by a 
smooth dependence.  
\end{abstract}
\date{\today}
\maketitle
\section{Introduction}

Since Lifshitz' seminal publication,\cite{Lifshitz60_1130}
the search for electronic topological transitions (ETT) of the Fermi surface
(FS)
has been a subject of permanent interest, for reviews see Refs.\
\onlinecite{BLANTER94_160} and \onlinecite{VARLAMOV89_469}. 
As a true phase transition it may
appear only at $T=0$ in highly pure samples, otherwise it is just a 
crossover.
While alloying may change the band filling and, thus, the number or
connectivity of the FS sheets, it unavoidably 
smears the electron states and the observable effects.
Magnetic field can cause ETT as well,\cite{Kozlova05_086403} but the 
field strength required to achieve a Lifshitz transition in a
conventional metal is beyond the present technical possibilities.
Thus, low-temperature pressure experiments are most promising
to find measurable effects related to ETT.\cite{BLANTER94_160}

All thermodynamic and transport properties of a metal are influenced by
the topology of the FS to some extend.
In the past, transport properties were in the focus of research,
since the impact of ETT on the transport is stronger than
on, e.g., elastic properties.\cite{VARLAMOV89_469}
As an example, Godwal {\em et al}.\ \cite{Godwal98_773} 
found a Lifshitz transition in AuIn$_2$, where measurements yield anomalies
in the electrical resistivity and in the thermoelectric power in the
$\unit[2-4]{GPa}$ pressure range, but angle-dispersive x-ray-diffraction 
measurements do not indicate any structural anomaly.
Very recently, an elastic anomaly was observed in YCo$_5$ under pressure, 
caused by an exceptionally strong van Hove singularity in the electronic
density of states (DOS).\cite{Rosner_submitted}

The question whether Lifshitz transitions
are visible in the
$c/a$ ratio of the hexagonal lattice parameters
of Zinc or if their effect is below the limit of
detectability is controversially discussed in the literature. 
Earlier experimental evidence\cite{Kenichi97_5170}
for an anomaly was later traced back to
non-hydrostatic pressure conditions,\cite{Kenichi99_6171}
and earlier theoretical confirmations of the elastic anomaly
were found to be caused by
insufficient ${\bf k}$-point sampling.\cite{Steinle-Neumann01_054103}
This discussion has recently been resumed\cite{Garg02_8795} by an alternative
presentation of the hydrostatic pressure data from 
Ref.\ \onlinecite{Kenichi99_6171}.

A similar puzzling situation is present in the case of the 5d element Osmium,
which is one of the densest elements and the metal with the largest
bulk modulus.
Occelli {\em et al.}\ 
measured the lattice parameters of Os under pressure,
found a discontinuity in the first pressure derivative of
the $c/a$ ratio at about 25 GPa, and assigned
this anomaly to a Lifshitz transition.\cite{Occelli04_095502}
On the other hand, Takemura did not infer
an anomaly from his related
data obtained by similar measurements.\cite{Kenichi04_012101}
Subsequently, two theoretical papers were published confirming
the existence of an anomaly
in $c/a$ versus pressure at about 10 GPa\cite{Sahu05_113106}
and at 27 GPa,\cite{Ma05_174103} respectively.
However, no reason for this anomaly was disclosed in the electronic structure.
In particular, the FS topology was found unchanged up to a
pressure of 80 GPa.\cite{Ma05_174103}

As an attempt to get a more detailed understanding of the dependence
of the $c/a$ ratio of Osmium under pressure, we have carried out
very precise density functional calculations applying the 
FPLO code,\cite{Koepernik99_1743} considering pressures up to 180 GPa. 
We identified three ETT in the high pressure region.
Further,
we confirm the finding of Ma and co-workers\cite{Ma05_174103}
that there is no ETT within the parameter range they considered.
Our calculations yield pressure dependent lattice parameters that 
are in accord with those presented by the previous authors.
Statistical analysis of both our and previously published
data eventually shows that there is no reason to assume a discernible
anomaly in the dependence
of $c/a$ on pressure from ETT. 

\section{Details of calculation}

Calculations were performed for the
hexagonal close-packed structure (hcp, P6$_3$/mmc, space group
194) which is the equilibrium structure of Osmium in the considered
pressure range.\cite{Hebbache06_6}
The four-component relativistic version of the 
full potential local orbital (FPLO)
band structure code,\cite{Koepernik99_1743}
release 5, was used. 
It accounts for kinematic relativistic effects including spin-orbit
coupling to all orders. In the case of Osmium, the $d_{3/2}$ - $d_{5/2}$
splitting is 1 eV. Therefore it cannot be neglected in the present
analysis.
After
test calculations with different basis sets we decided to use a minimum basis
consisting of $4f5s5p$ states in the semicore and $6s6p5d$ states in the
valence. All lower lying states were treated as core states. 
A ${\bf k}$-mesh subdivision
of $48\times48\times48$ ${\bf k}$-points in the 
full Brillouin zone (BZ) was used, which
corresponds to $5425$ ${\bf k}$-points in the irreducible part of BZ.
This mesh is significantly finer than the meshes used by Ma {\em et al.},
$16\times16\times12$ in the BZ,\cite{Ma05_174103}
and by Sahu and Kleinman, $84$ points\cite{Sahu05_113106} in the 
irreducible part of BZ.
For calculating the density of states, the ${\bf k}$-mesh was refined to 
$96\times96\times96$ ${\bf k}$-points in the full BZ ($40033$ ${\bf k}$-points
in the irreducible BZ).
The ${\bf k}$-space integrations were carried out with the linear
tetrahedron method.\cite{LEHMANN72_469}
For the exchange-correlation potential we employed the
local density approximation (LDA) in the version proposed by 
Perdew and Wang, 1992.\cite{PERDEW92_13244}

\section{Total energy calculations}
\begin{figure}[t!]
\BoxedEPSF{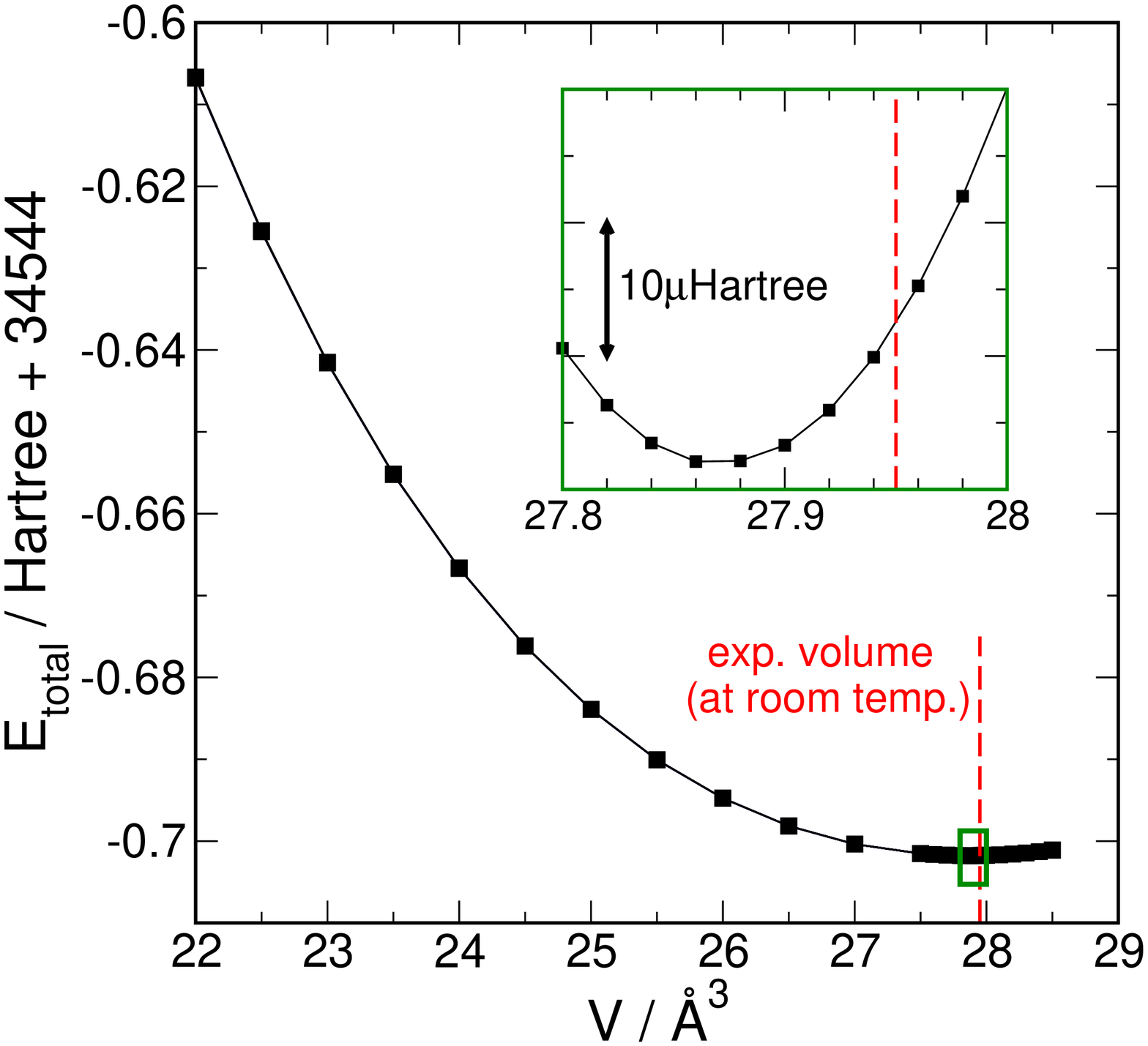 scaled 250}
\BoxedEPSF{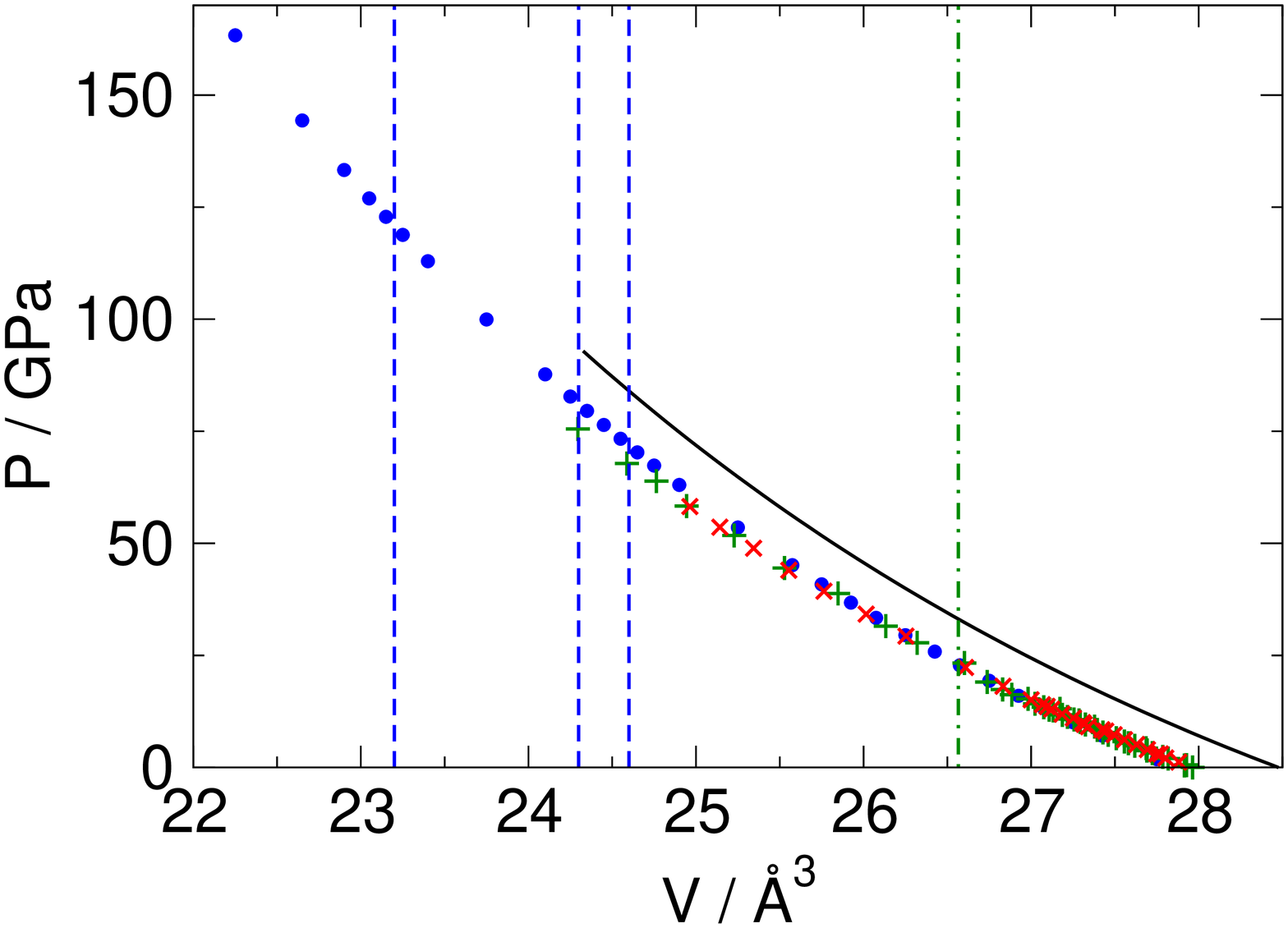 scaled 250}
\caption{(color online). Upper panel: total energy versus unit cell 
	volume with relaxed $c/a$ ratio.
	The red (gray) dashed line marks the 
	experimental room-temperature volume.
	Inset: magnification of the total energy versus volume curve
	around the minimum. In the large graph this region is marked with
	a green (gray) box. Lower panel:
	equation of state of Osmium. Blue (black) circles:
        present calculations; green (gray)
        pluses and red (gray) crosses: experimental data from
        Occelli {\em et al.}\ \cite{Occelli04_095502} 
	and from Takemura,
        \cite{Kenichi04_012101} respectively; black line: 
        calculations by Ma {\em et al.}\ \cite{Ma05_174103}.
        The green (gray) dashed-dotted line denotes the volume
        where Occelli {\em et al.}\ \cite{Occelli04_095502} 
        find an elastic anomaly. 
        The blue (black) dashed lines indicate the volumes where
        Lifshitz transitions are found in the present calculations.
        }\label{OsEPV}
\end{figure}
The upper panel of Figure \ref{OsEPV} shows the total energy 
of Osmium plotted against
volume $V$, where the $c/a$ ratio has been relaxed 
for each volume, $E_{\rm total}(V) = \min_{(c/a)} E_{\rm total}(V, c/a)$,
which is the correct condition for hydrostatic pressure. 
The red (gray) dashed line shows the 
experimental volume (taken from Ref.\ \onlinecite{Occelli04_095502}), 
determined in a room-temperature experiment. 
The calculated equilibrium volume is $0.3\%$ smaller
than the measured room-temperature volume. Accounting for thermal expansion
provides an almost perfect agreement.
It is a known but not yet
understood feature of LDA, that it reproduces the ground state volumes
of the heavy 5d elements much better than gradient
approximations.\cite{chapterHRO}
We take the agreement in the present case
as justification to rely on LDA in the
further investigations.

The inset of Figure \ref{OsEPV}, upper panel, proves the exceptional stability
of the calculations. Each data point in the figure is taken from 
independent self-consistency runs and independent $c/a$
optimizations.
The FPLO code \cite{Koepernik99_1743} yields a very smooth energy-curve,
with numerical noise well below the $\mu$Hartree range. 
We took advantage of this stability in the evaluation of the
equation of state (EOS) and the pressure dependence of $c/a$, presented
in Section \ref{sec_geom}:
While the calculations yield directly the two dependences $E_{total}(V)$
and $(c/a)(V)$, in experiment the pressure $P=-(dE_{total}/dV)$
may be measured independently.
Instead of relying on an analytic fit of Birch-Murnaghan type 
for $E_{total}(V)$ we 
calculated the derivative numerically by means of the three-point formula,
$f'((x_1+x_2)/2)\approx (f(x_2)-f(x_1))/(x_2-x_1)$, considering neighboring 
$V$ values.
This seemingly less sophisticated procedure solves a dilemma, which is
present for analytic fits. If an analytic fit is used for the
whole volume range considered, the anomaly will be removed by the fit.
If the analytic fit is done piece-wise, anomalies will necessarily be
produced at the end points of the respective fit regions, due to 
different noise in the adjacent regions.
Previous authors used one fit for the whole 
range.\cite{Sahu05_113106,Ma05_174103}
We checked our procedure by comparing the evaluated pressures with 
five different fits of Birch-Murnaghan type. For 80\% of the pressure
range between zero and 180 GPa, our numerical fit lies within the
$P(V)$ values spanned by the different analytic fits. The maximum
difference between any of the analytic fits and the numerical fit
amounts to $0.2$ GPa, and the mean difference amounts to $0.1$ GPa.

Figure \ref{OsEPV}, lower panel, shows the EOS of Osmium.
Green (gray) pluses and red (gray) crosses denote 
experimental data by Occelli {\em et al.}\ 
\cite{Occelli04_095502} and by Takemura,\cite{Kenichi04_012101} 
respectively. Both data sets do not deviate from each other on the
scale of this figure.
The blue (black) circles denote
results of the present calculations,
obtained from the $E(V)$ data, upper panel of Figure \ref{OsEPV},
by numerical differentiation.
The EOS calculated using LDA coincides with the experimental data
in the low-pressure range up to about 40 GPa. It slightly over-estimates
the pressure in the higher-pressure range.

Results from calculations by Ma {\em et al.}\ \cite{Ma05_174103}
are depicted by a black line.
These calculations were carried out in the 
GGA which systematically over-estimates the zero-pressure
volume of heavy 5d metals.\cite{chapterHRO}
Accordingly, the GGA data run almost parallel to the LDA data, with 
a volume offset of about 2\%.
An alternative presentation, $P(V/V_0)$ with $V_0 = V(P=0)$,
yields a very good coincidence of both calculated data sets.
As stated before,
the offset explains the fact that Ma {\em et al.} did not find
any ETT in their calculations up to 80 GPa,\cite{Ma05_174103} 
while we find the first
ETT at about 72 GPa.
 
In the lower panel of Fig.\ \ref{OsEPV}, there is no anomaly of the 
EOS visible in any data set at any pressure, in
agreement with the earlier presentations of the quoted data and with
the weakness of the Lifshitz transitions disclosed in this paper.

\section{The Fermi surface under pressure}\label{FermiS}
It was suggested by Occelli {\em et al.}, that the observed
anomaly in $(c/a)(P)$ should arise from an ETT.
Thus, we first investigate the volume dependence of the
FS in the range $\unit[22.0]{\AA^3}\leq V \leq \unit[28.0]{\AA^3}$.
This volume dependence translates into a pressure
dependence through the EOS given in the lower panel of 
Figure \ref{OsEPV}, 
$P \lesssim 180$ GPa. 
As in the evaluation of the total energy, $c/a$ has been relaxed for
each volume. 

\begin{figure}
\BoxedEPSF{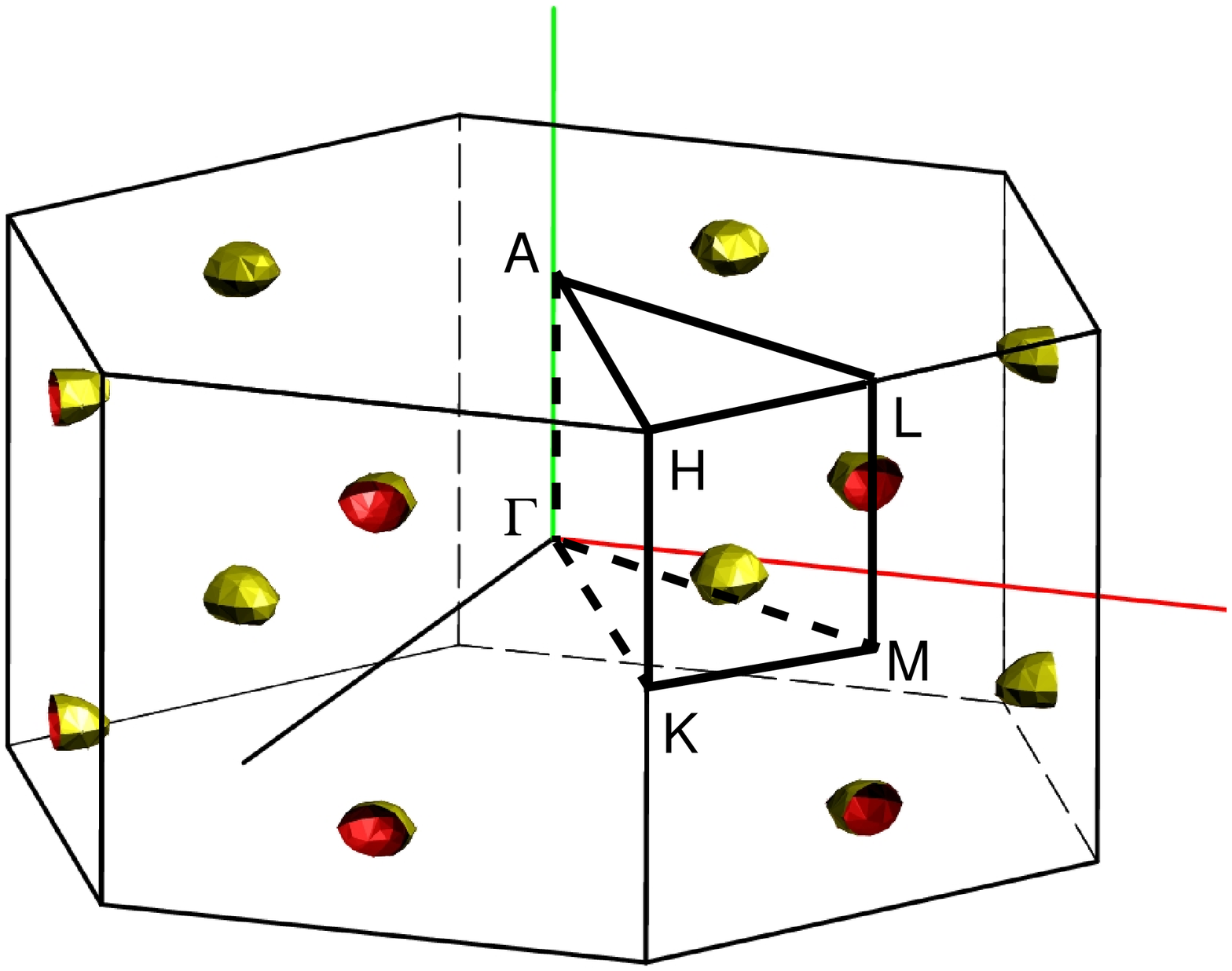 scaled 200}
\BoxedEPSF{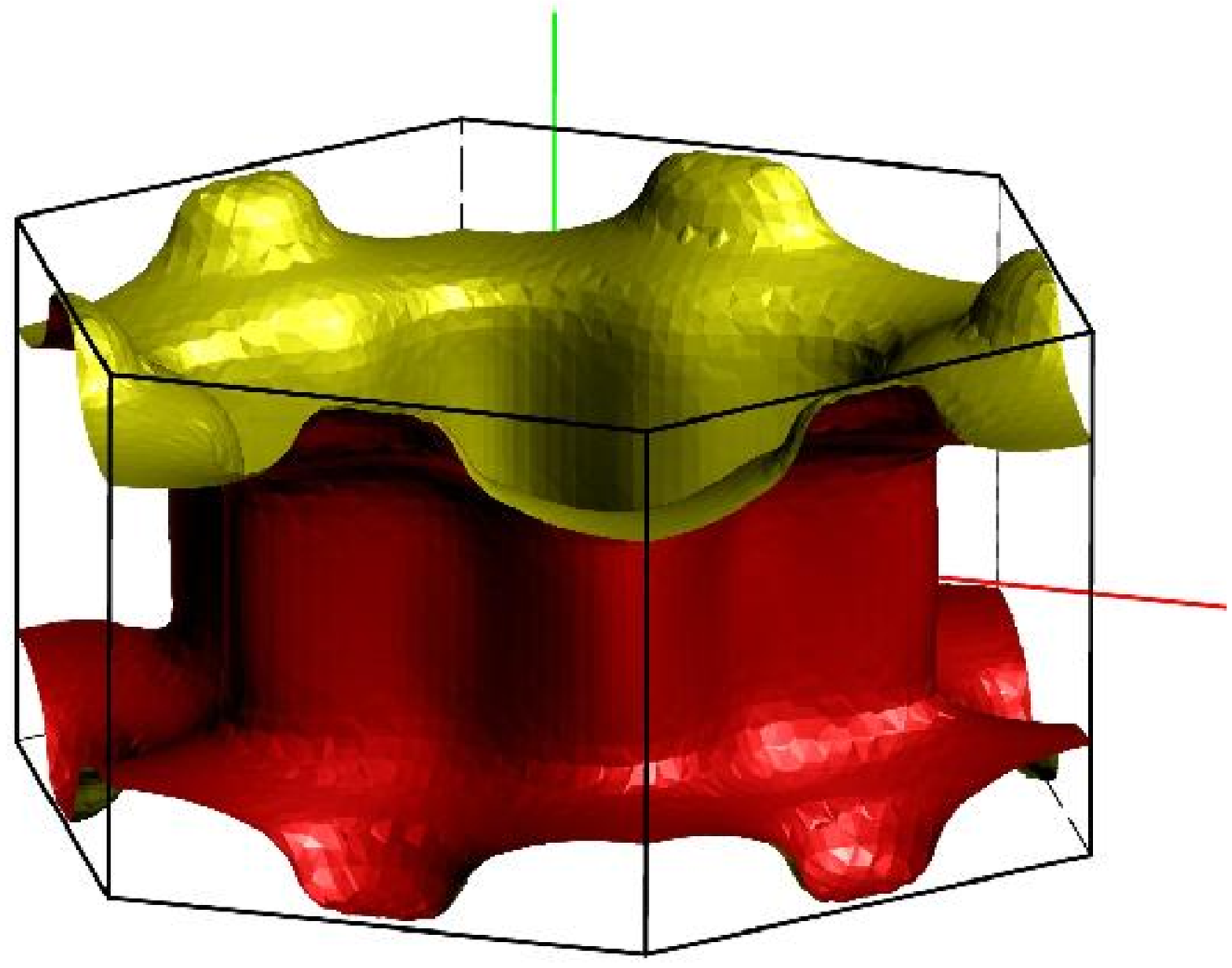 scaled 200}
\BoxedEPSF{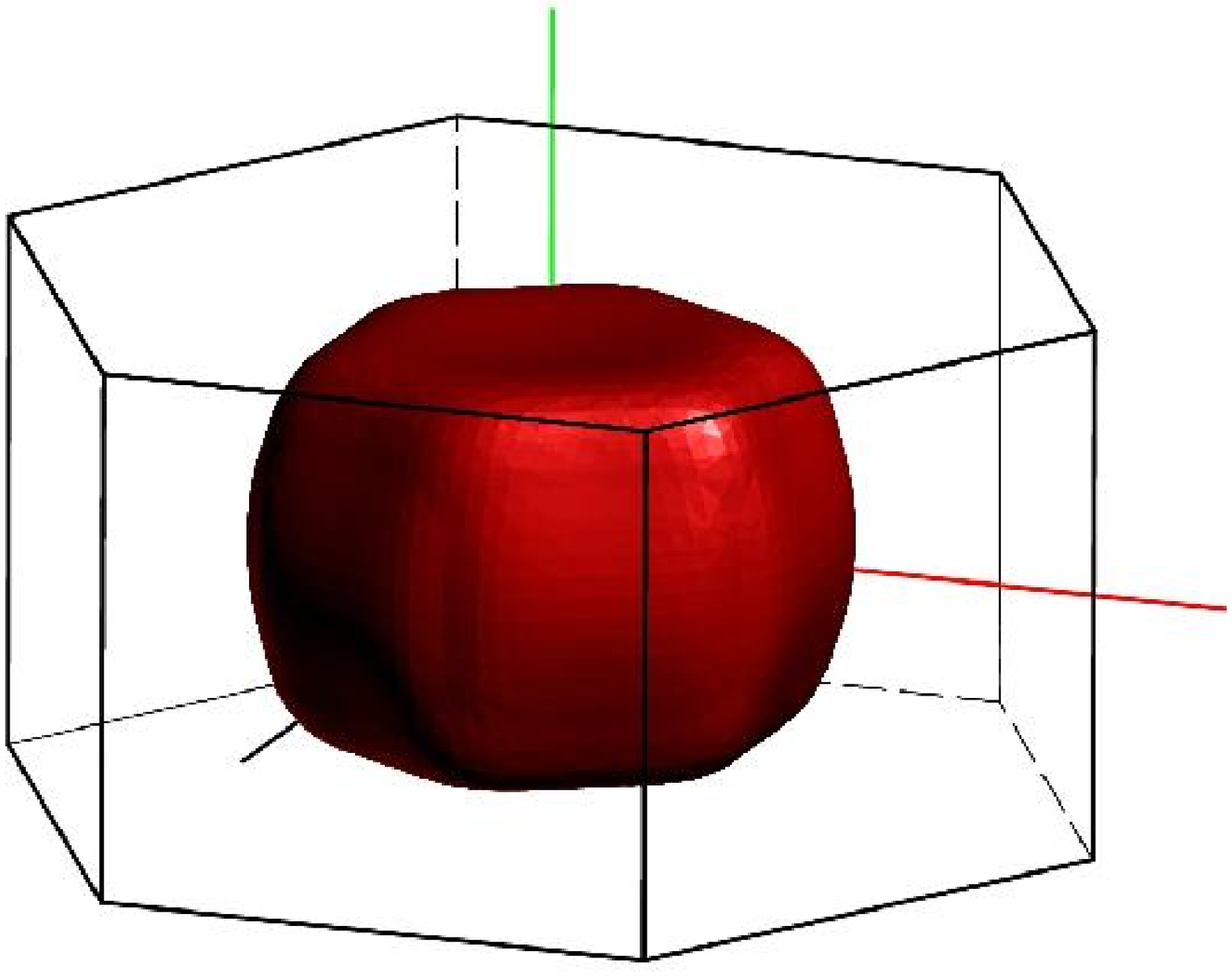 scaled 200}
\BoxedEPSF{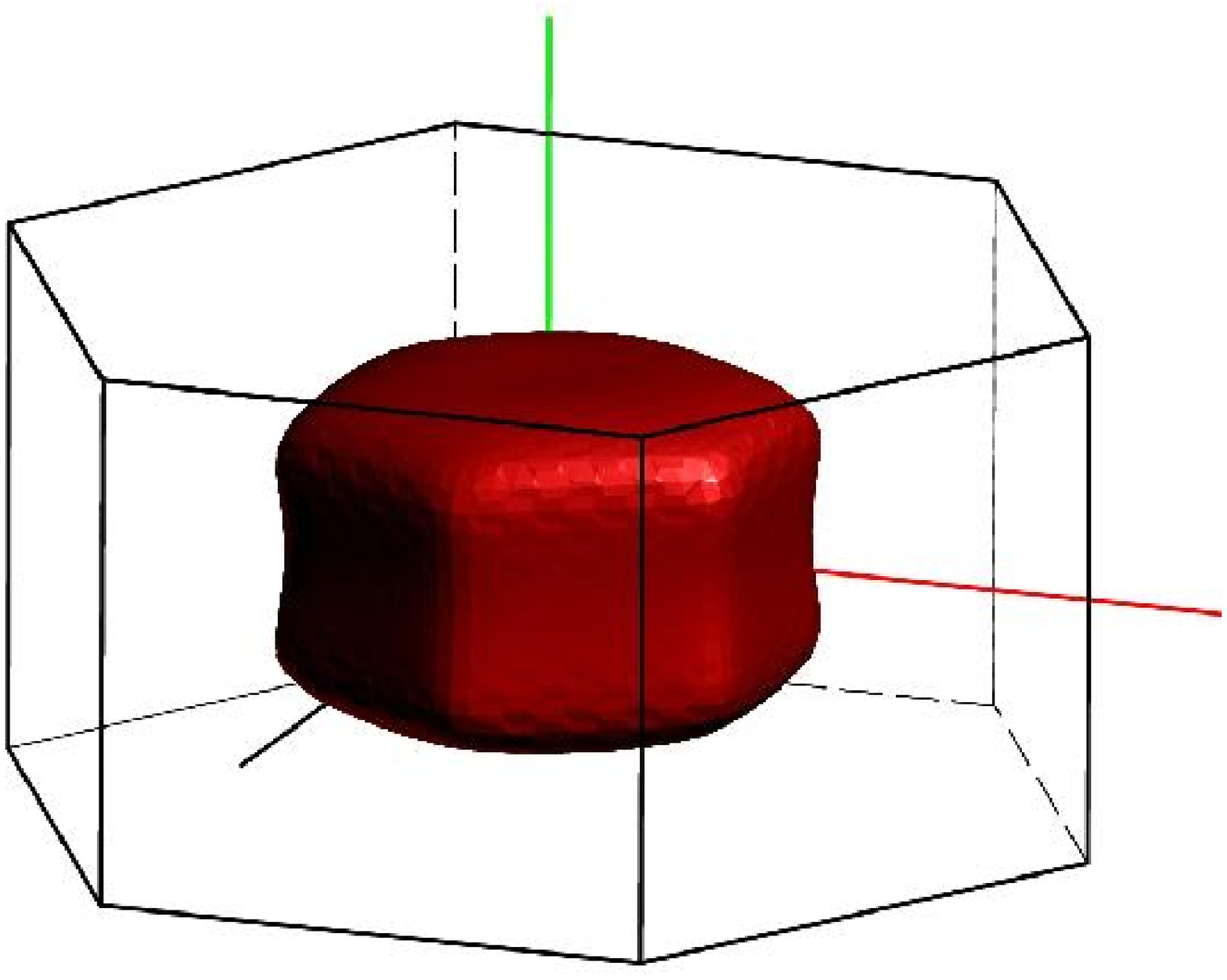 scaled 200}
\caption{(color online). The Fermi surface of 
	Osmium at zero pressure. First, second, third,
and fourth sheets are given separately in the upper left, upper right,
lower left, and lower right panel, respectively. Red (black) and yellow 
(light-gray) colors
denote occupied and unoccupied sides of the Fermi surface, respectively.}
\label{FSeq}
\end{figure}

\begin{figure}[t!]
\BoxedEPSF{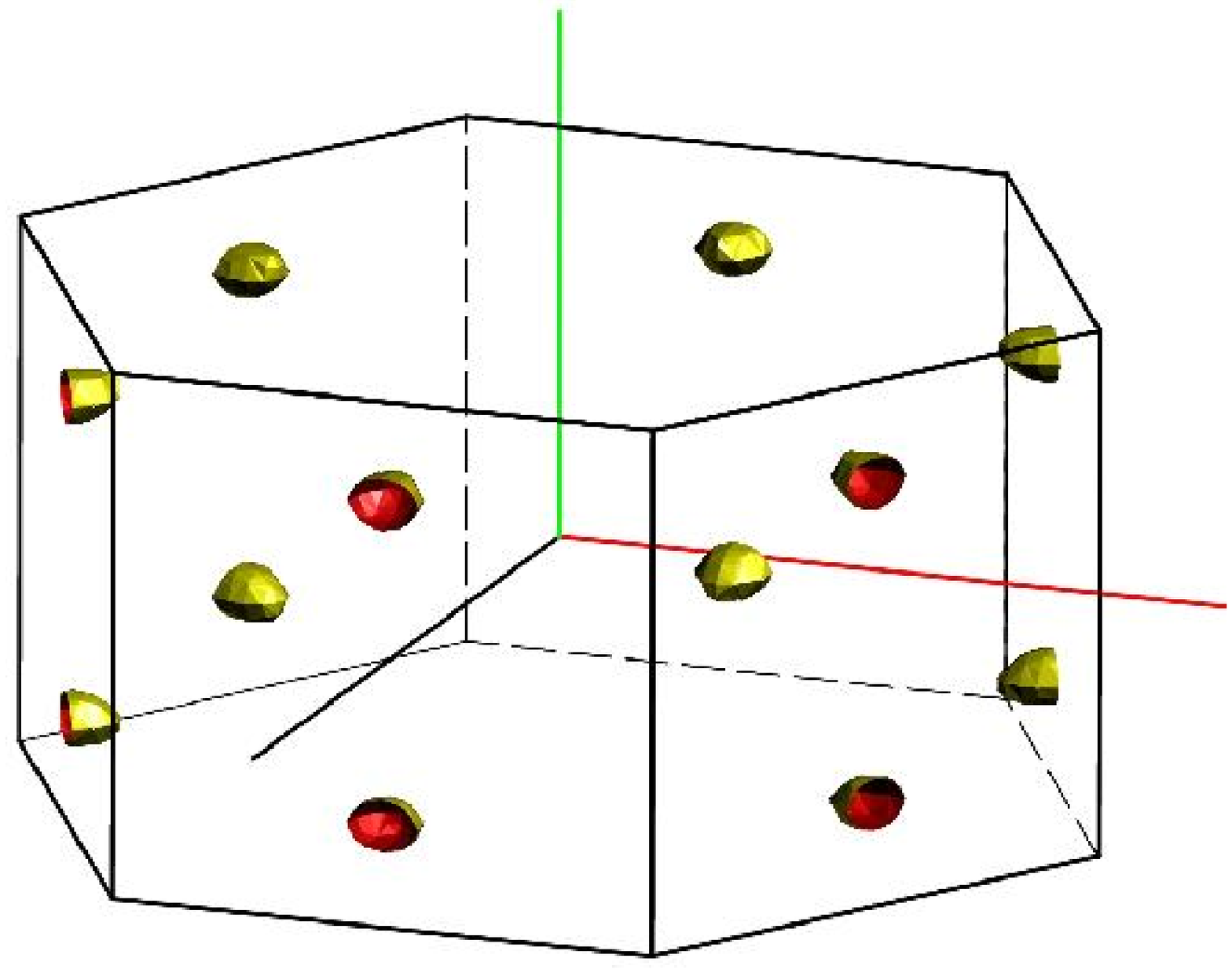 scaled 200}
\BoxedEPSF{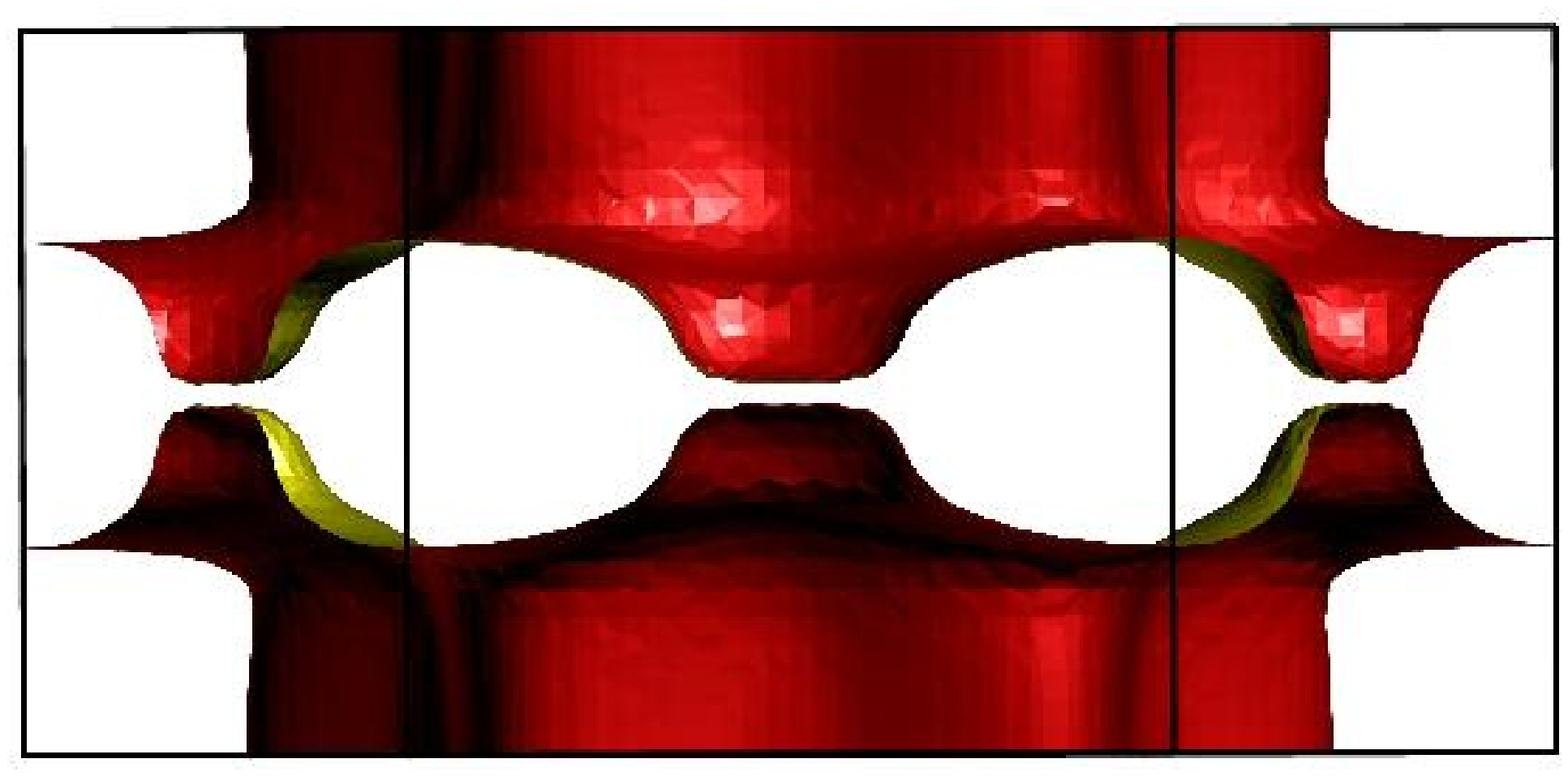 scaled 200}
\BoxedEPSF{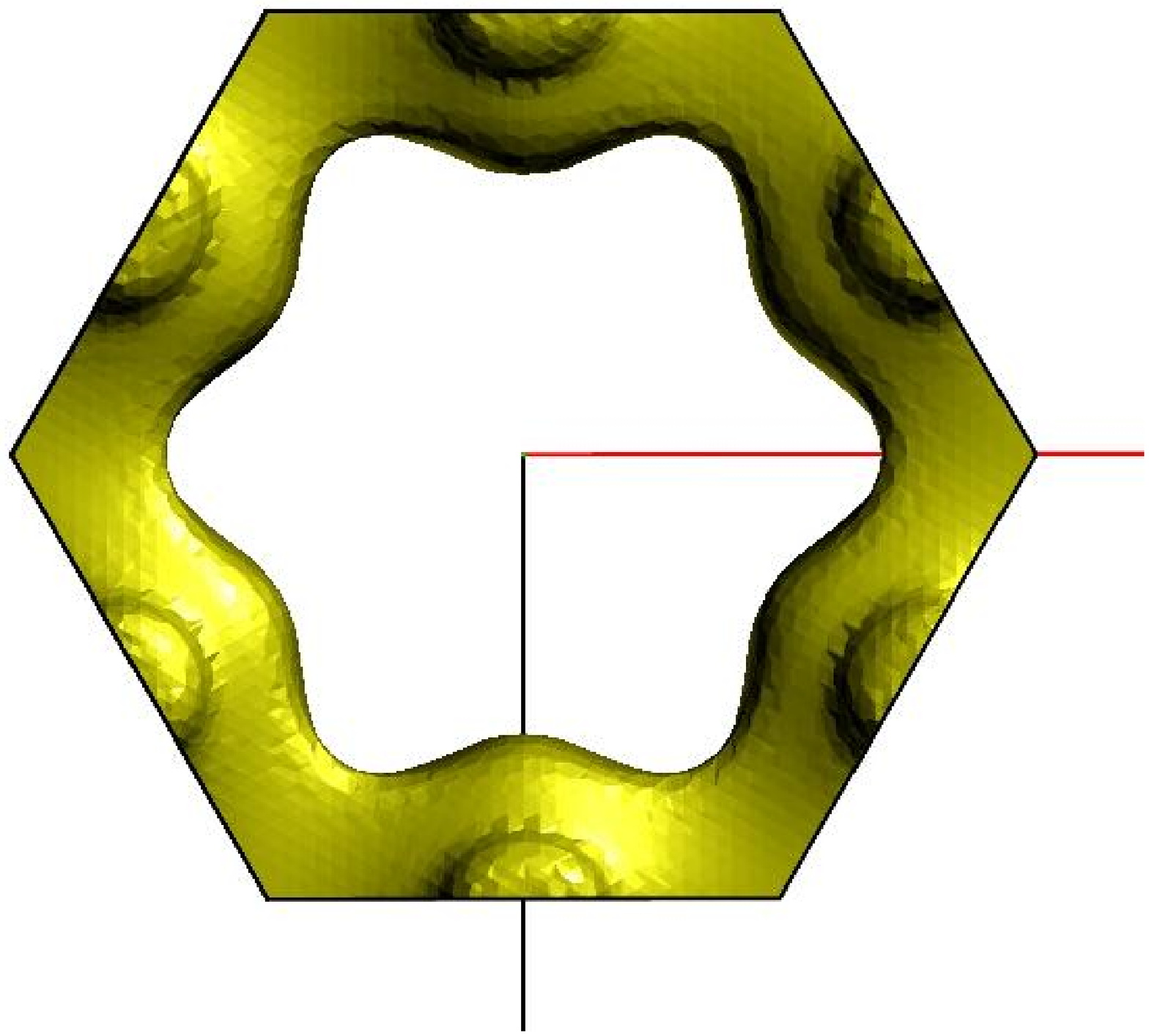 scaled 200}
\BoxedEPSF{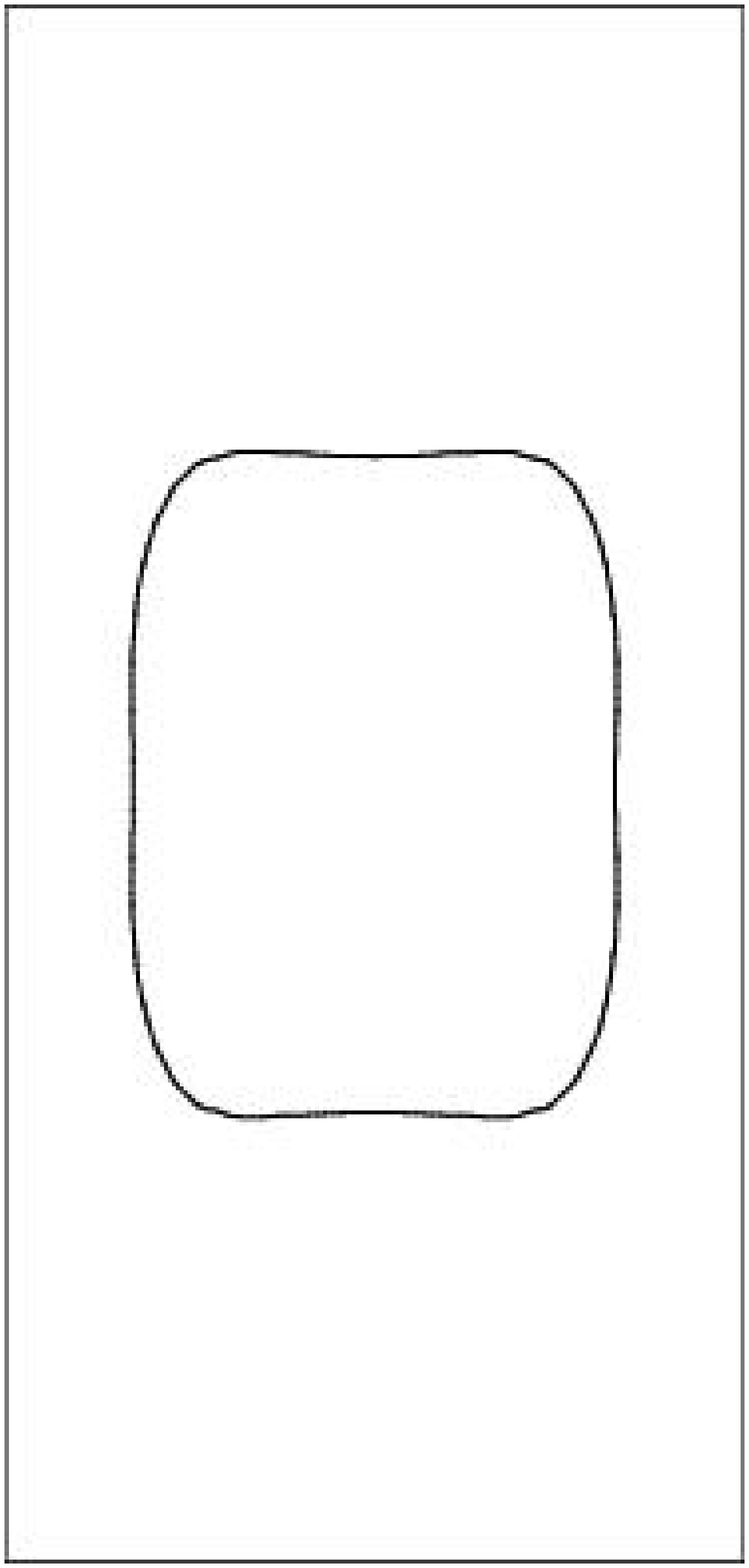 scaled 250}\\[10mm]
\BoxedEPSF{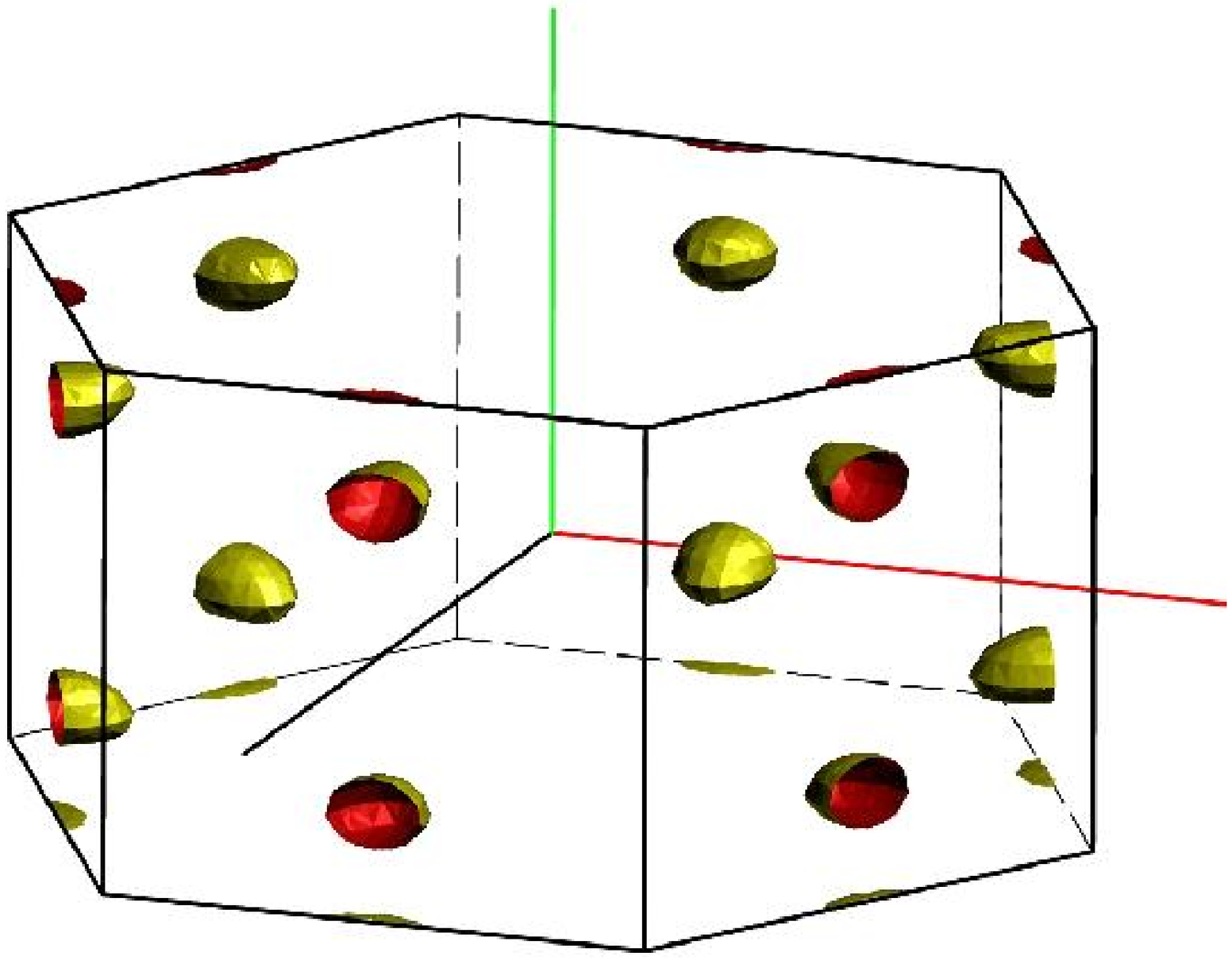 scaled 200}
\hspace{-0.13cm}
\BoxedEPSF{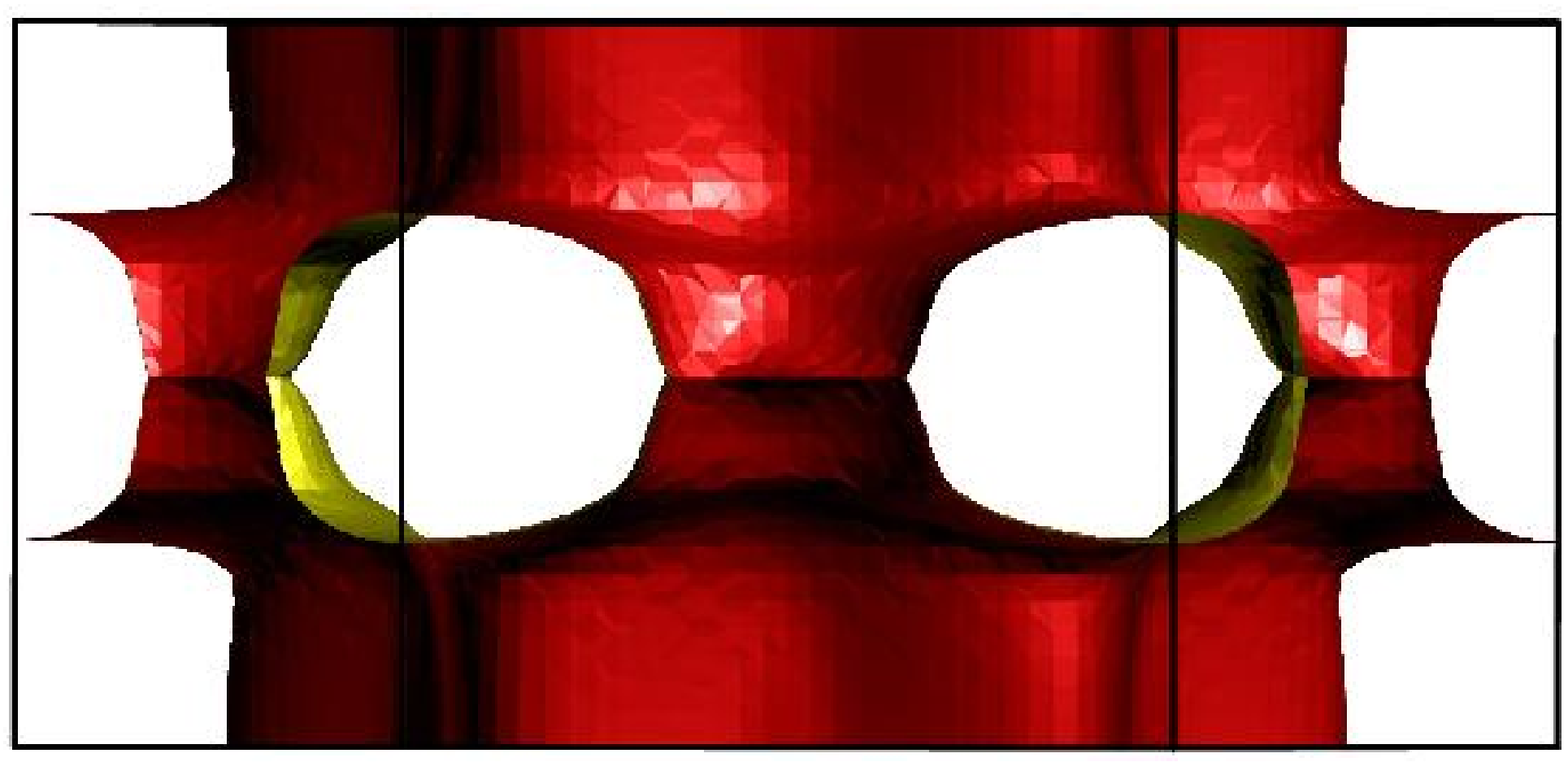 scaled 200}
\hspace{-0.13cm}
\BoxedEPSF{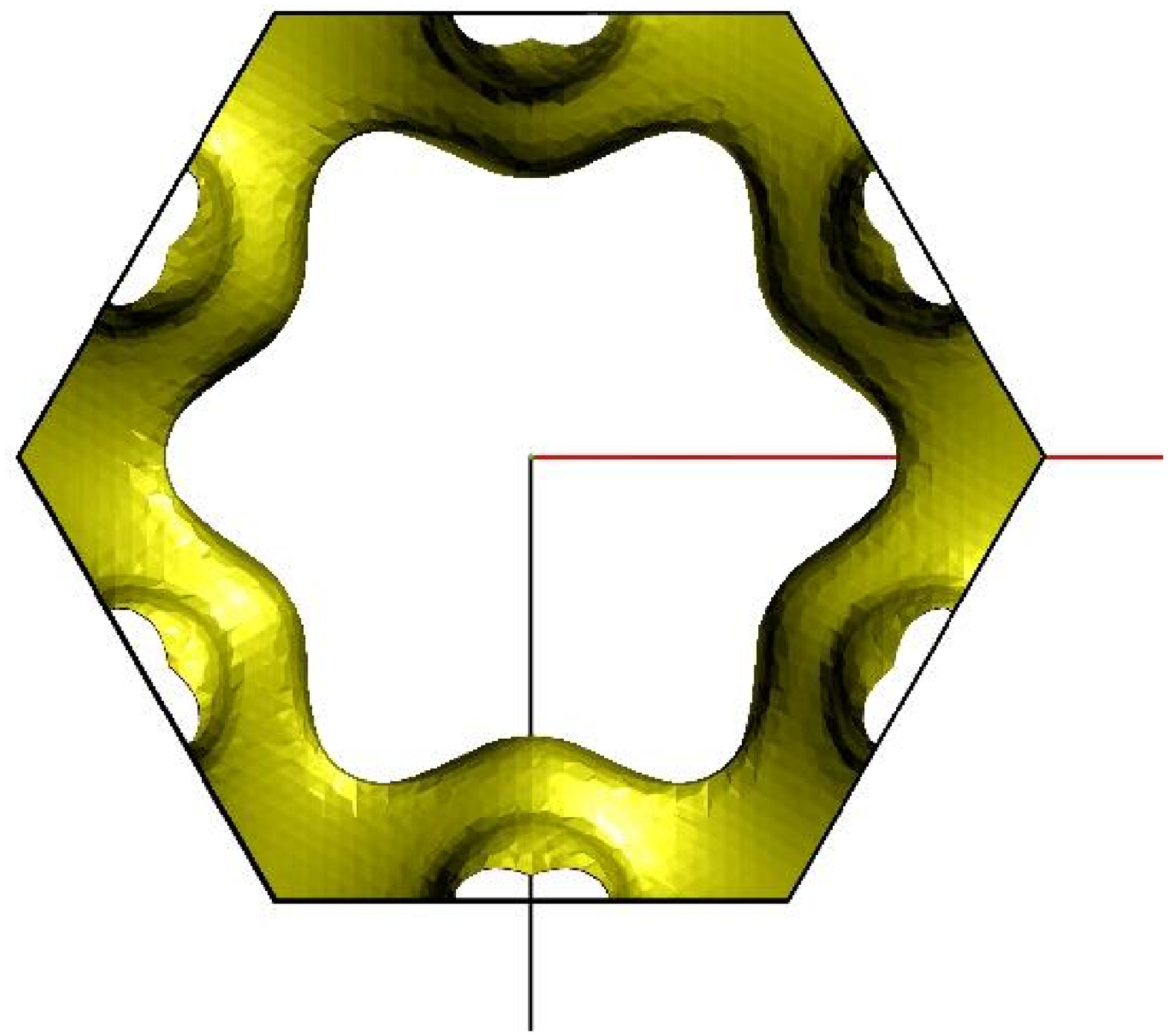 scaled 200}
\BoxedEPSF{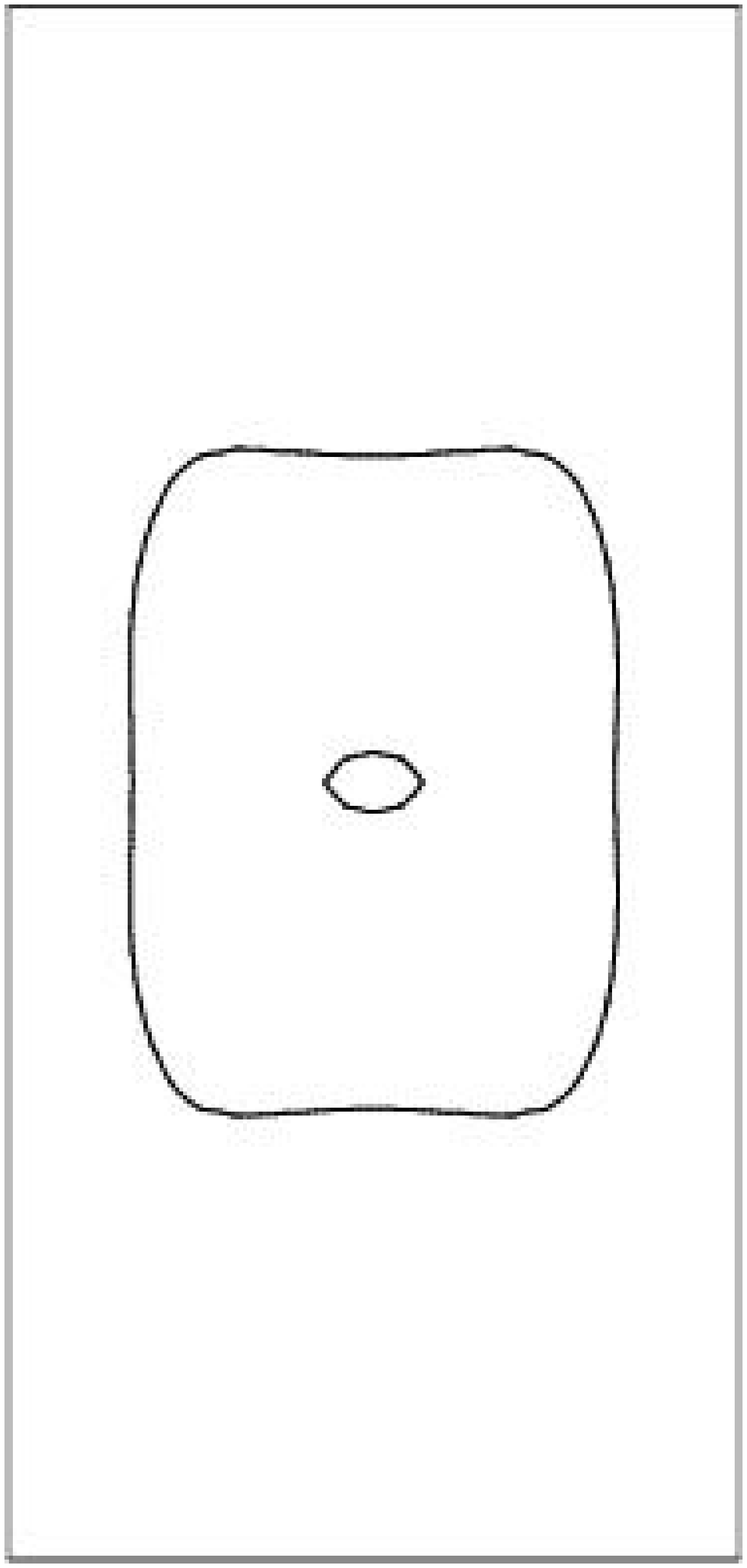 scaled 250}
\caption{(color online). Comparison between the Fermi surface 
	of Osmium at zero pressure
($V=\unit[27.87]{\AA^3}$ first row) and under high pressure 
(second row, $P\approx\unit[180]{GPa}$, $V=\unit[22.00]{\AA^3}$).
The upper left panels show the first FS sheet, 
the upper right and the lower left panels show the second FS sheet 
in a side view and in a top view, respectively.
       In order to make the connectivity better visible,
       the side view is shifted in $z$-direction, such
that the BZ edge lies in the middle of the picture.
The lower right panels display a cut
through the fourth sheet of the Fermi surface in the y-z plane.
	Colors are explained in Figure \ref{FSeq}.}
\label{FSvergleich}
\end{figure}

At zero pressure, the FS of Osmium
has four sheets, see Figure \ref{FSeq}.
The first sheet consists of 
small hole ellipsoids located 
at the line L-M.
Occelli {\em et al.} suspected these ellipsoids to disappear
under pressure.\cite{Occelli04_095502}
The second sheet is
a big monster surface, which is closed in z-direction (compare
Figure \ref{FSvergleich})
but connected in the x-y plane. The third and fourth sheets are
closed surfaces, which are
nested and centered around the $\Gamma$-point.  The 
inner one of the two is waisted. 
Our calculated zero-pressure FS agrees well
with measurements of Kamm and Anderson,\cite{Kamm70_2944}
with
the calculated FS of Smelyansky {\em et al.},\cite{Smelyansky90_9373} and
with the more recent calculation of Ma {\em et al.}.\cite{Ma05_174103}

Figure \ref{FSvergleich} shows a comparison of the FS
at zero pressure (first row, $V=\unit[27.87]{\AA^3}$, $c/a=1.5850$) 
with
the FS calculated at a very high pressure of about 
$\unit[180]{GPa}$ (second row, $V=\unit[22.0]{\AA^3}$, $c/a=1.5968$). 
It is obvious, that the small hole ellipsoid between L and M
does not disappear under pressure. On the contrary, it grows.
Nevertheless, we find a Lifshitz transition in the first FS sheet:
a further tiny hole ellipsoid appears under pressure at the L-point. 
Another Lifshitz transition takes place in the second FS.
Here, a neck is created under pressure, 
centered at L.
The third FS sheet is not displayed in Figure \ref{FSvergleich} because
its topology is preserved in the considered pressure range.
Finally, a third ETT is found inside the fourth FS sheet,
where a hole ellipsoid appears under pressure at the $\Gamma$-point.

\begin{figure}
\BoxedEPSF{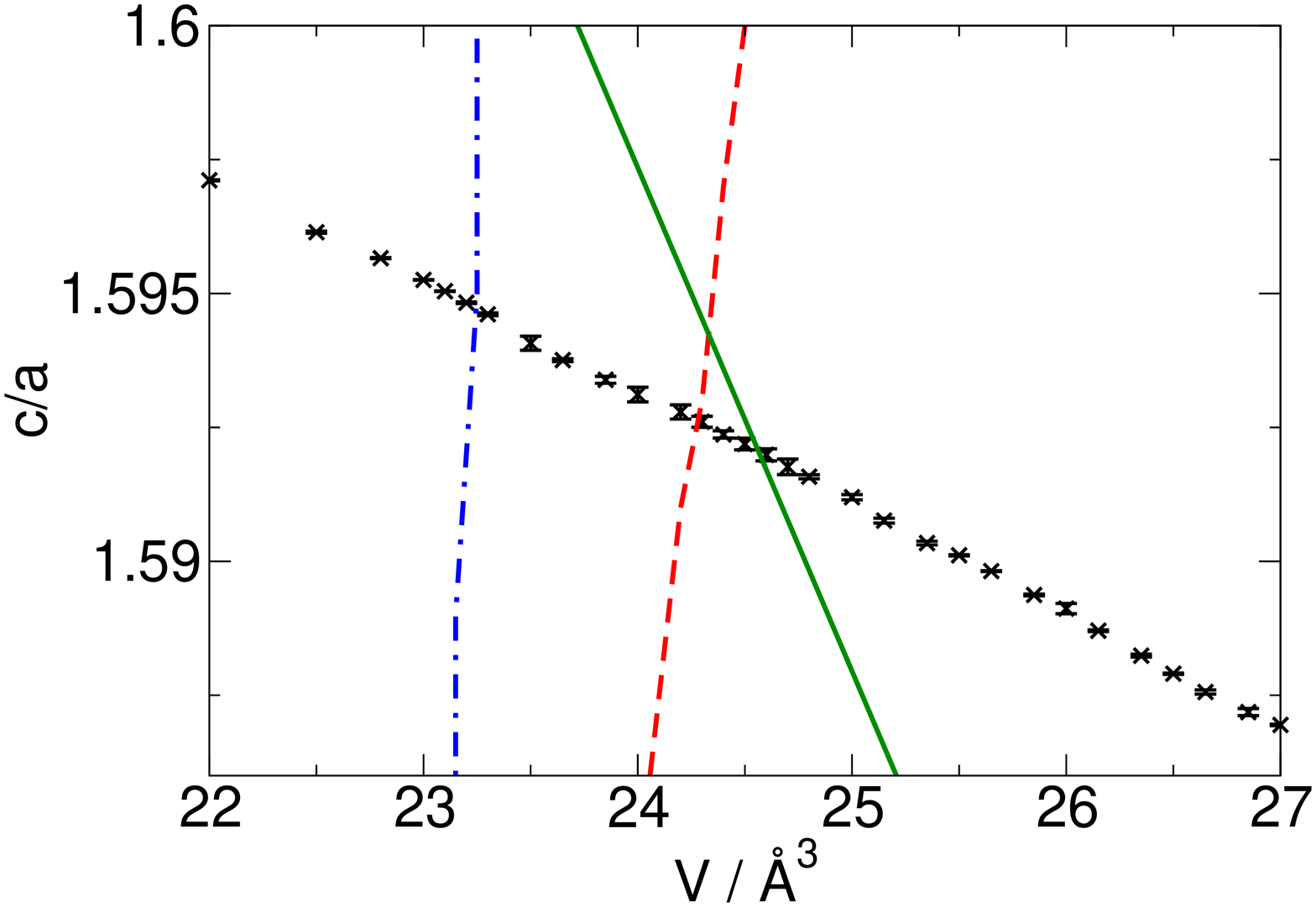 scaled 250}
\caption{(color online). Dependence of the FS topology on volume and $c/a$.
	Black symbols with error bars: calculated relaxed $c/a$ versus 
	unit cell volume. Green (gray) full line:
        ETT at the $\Gamma$-point. Red (gray) dashed line: 
	ETT at the symmetry-line LH.
        Blue (black) dash-dotted line: ETT at the L-point.}
\label{OsETT}
\end{figure}

\begin{figure}
\BoxedEPSF{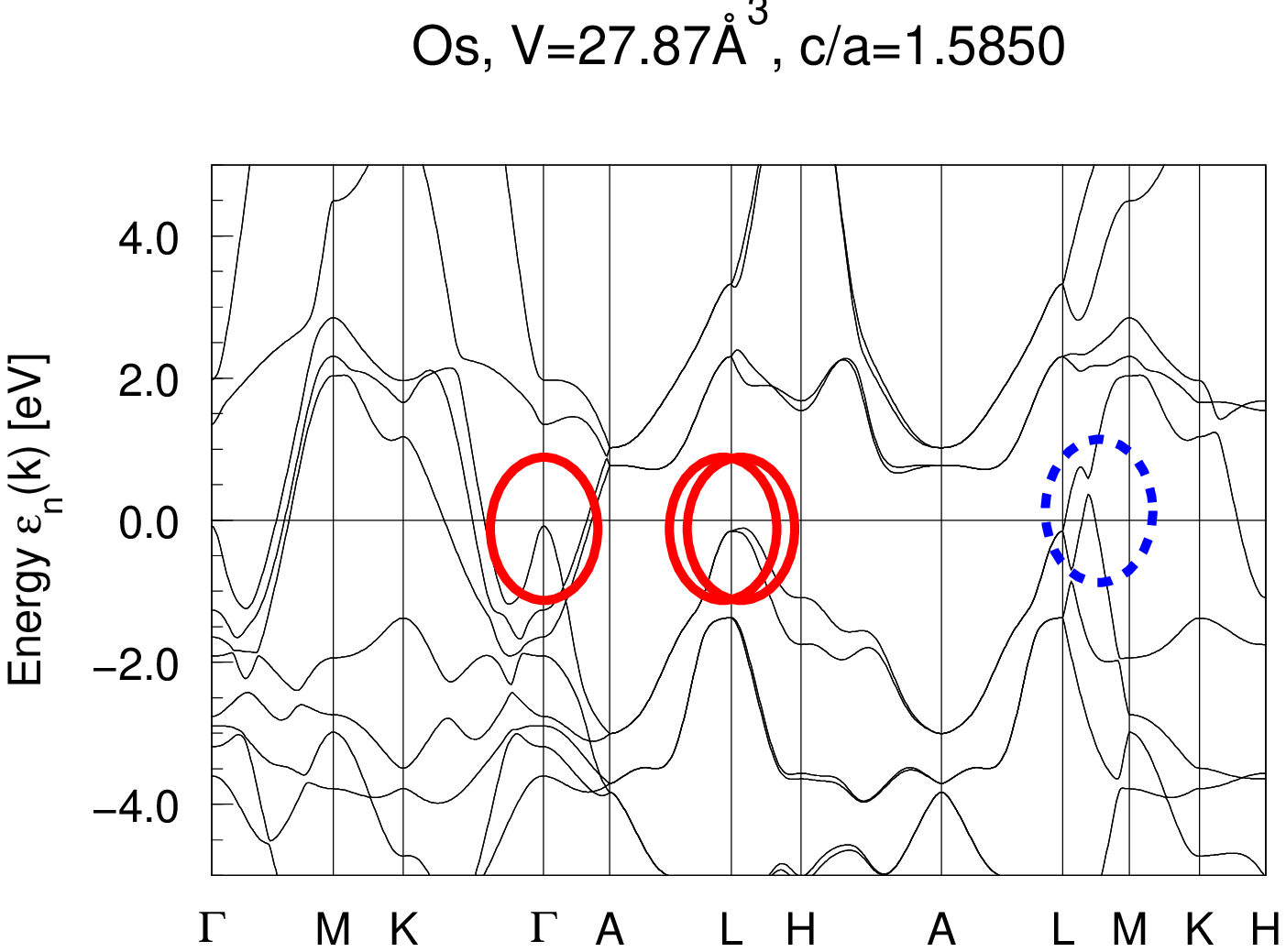 scaled 470}\\[5mm]
\BoxedEPSF{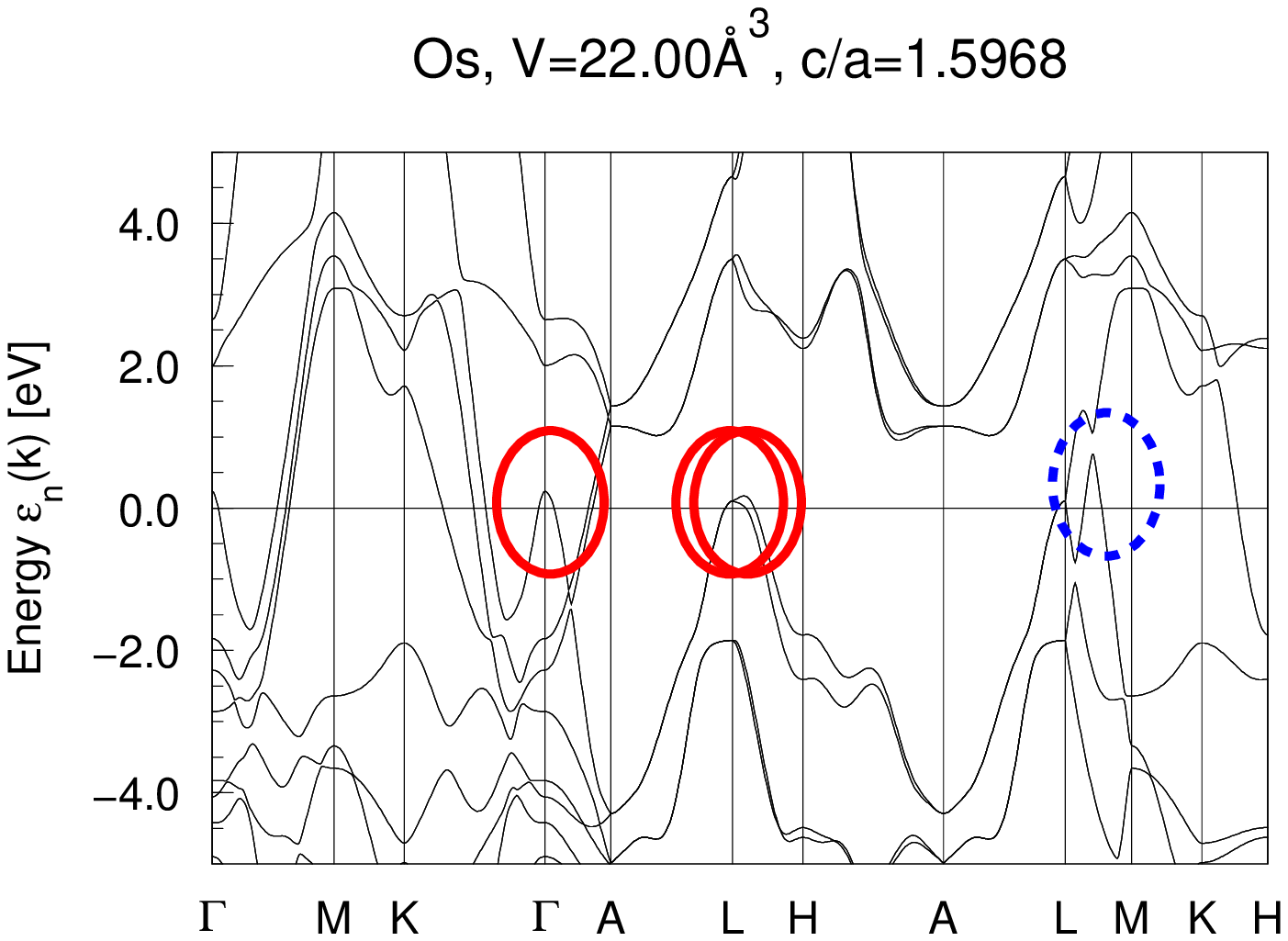 scaled 470}
\caption{(color online). Band structure of Osmium. Upper panel: zero pressure;
	lower panel: $P\approx\unit[180]{GPa}$.
	The red (gray) circles (full lines)
        mark points in ${\bf k}$-space where a band maximum 
	crosses the Fermi level under
        pressure. The blue (black) circle (dashed lines) 
	marks the location of an ETT
        proposed by Occelli {\em et al.}.\cite{Occelli04_095502}
        }\label{bandstrOs}
\end{figure}

At which pressure/volume the FS changes its topology depends on the 
$c/a$ ratio. This can be seen in Figure \ref{OsETT}.
Reducing the volume under hydrostatic pressure, 
starting at the zero-pressure value, one finds the
following sequence of transitions:
At first, the additional ellipsoid at the $\Gamma$-point appears at
$V=\unit[24.60]{\AA^3}$,
then the neck is formed at
$V=\unit[24.20]{\AA^3}$, and
finally the additional ellipsoid at the L-point appears at
$V=\unit[23.20]{\AA^3}$.
The related transition pressures are 72 GPa, 81 GPa, and 122 GPa.
Worth mentioning, a peculiarity relates the latter two ETT.
Namely, two necks are formed at first
at the mentioned volume which are situated
at the line L-H. These necks merge at L at the same pressure that lets the
L-point ellipsoid appear. This coincidence is due to a degeneracy
of the first and second sheet bands along L-A, see
Figure \ref{bandstrOs}.

If we presume the experimentally measured and extrapolated 
$(c/a)(V)$, the ETT take place
at pressures of about 50 GPa (ellipsoid at the $\Gamma$-point), at about
87 GPa (neck at the line L-H) and at about 122 GPa (ellipsoid at the L-point).

The discussed topological changes can be clearly identified in 
the band structure plots.
For each hole ellipsoid appearing under pressure one expects a related
maximum in the band dispersion to cross the Fermi energy.
In the case of the neck, a saddle-point of band dispersion 
should cross the Fermi level. This saddle-point does not lie on a 
symmetry-point but only on a symmetry-line of the Brillouin zone and thus 
appears only as a maximum in the band structure plot.

Figure \ref{bandstrOs} shows the band structure for zero pressure
(upper panel) and under a high hydrostatic pressure of about $\unit[180]{GPa}$
(lower panel). 
The loci of the three discussed ETT are marked by red (gray) full-line circles.
The pressure-induced Fermi level crossing of all these three band
maxima is obvious.
The blue (black) dashed-line circles in Figure \ref{bandstrOs} mark the place 
of the disappearance of a hole pocket proposed in Ref.\
\onlinecite{Occelli04_095502}.
The related maximum does, however, not fall below the Fermi level
under pressure but gets shifted to higher energy instead.

Summarizing this section, we find three distinct Lifshitz transitions
in the pressure range between 70 GPa and 130 GPa.
In the range of 25 GPa the present calculations do not yield an anomaly
of $c/a$ caused by the electronic structure.
None of the transitions disclosed here were found in 
the earlier calculations restricted to a
pressure range just below the first observed transition.\cite{Ma05_174103}
Note, that the Fermi surface topology in the latter publication was
investigated up to 80 GPa. The related {\em volume}, however, was slightly
larger than our predicted first critical volume.
The reason is that Ma {\em et al.} used the generalized gradient approximation
(GGA), while we used LDA.

\section{Density of states}

\begin{figure}[t!]
\BoxedEPSF{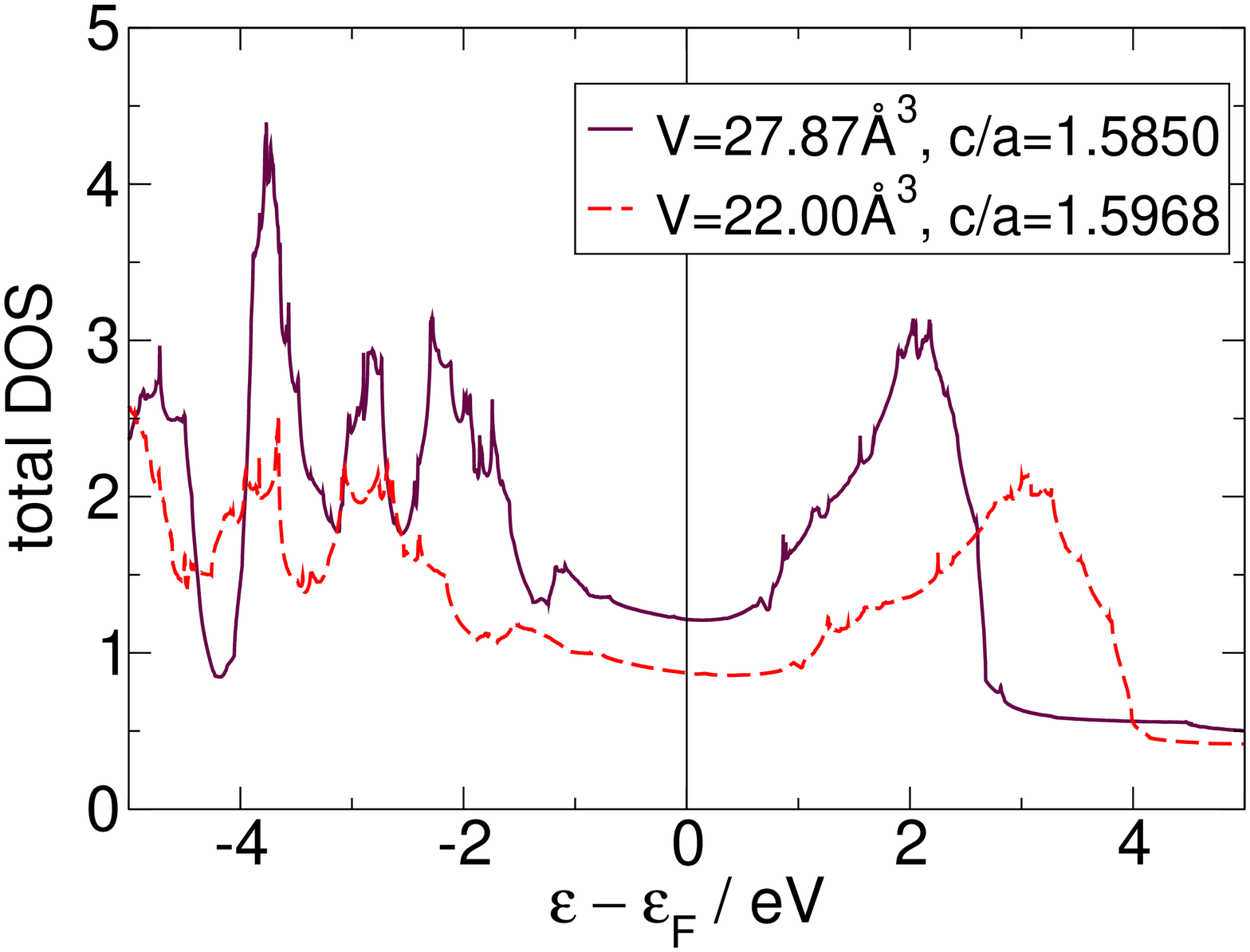 scaled 280}
\BoxedEPSF{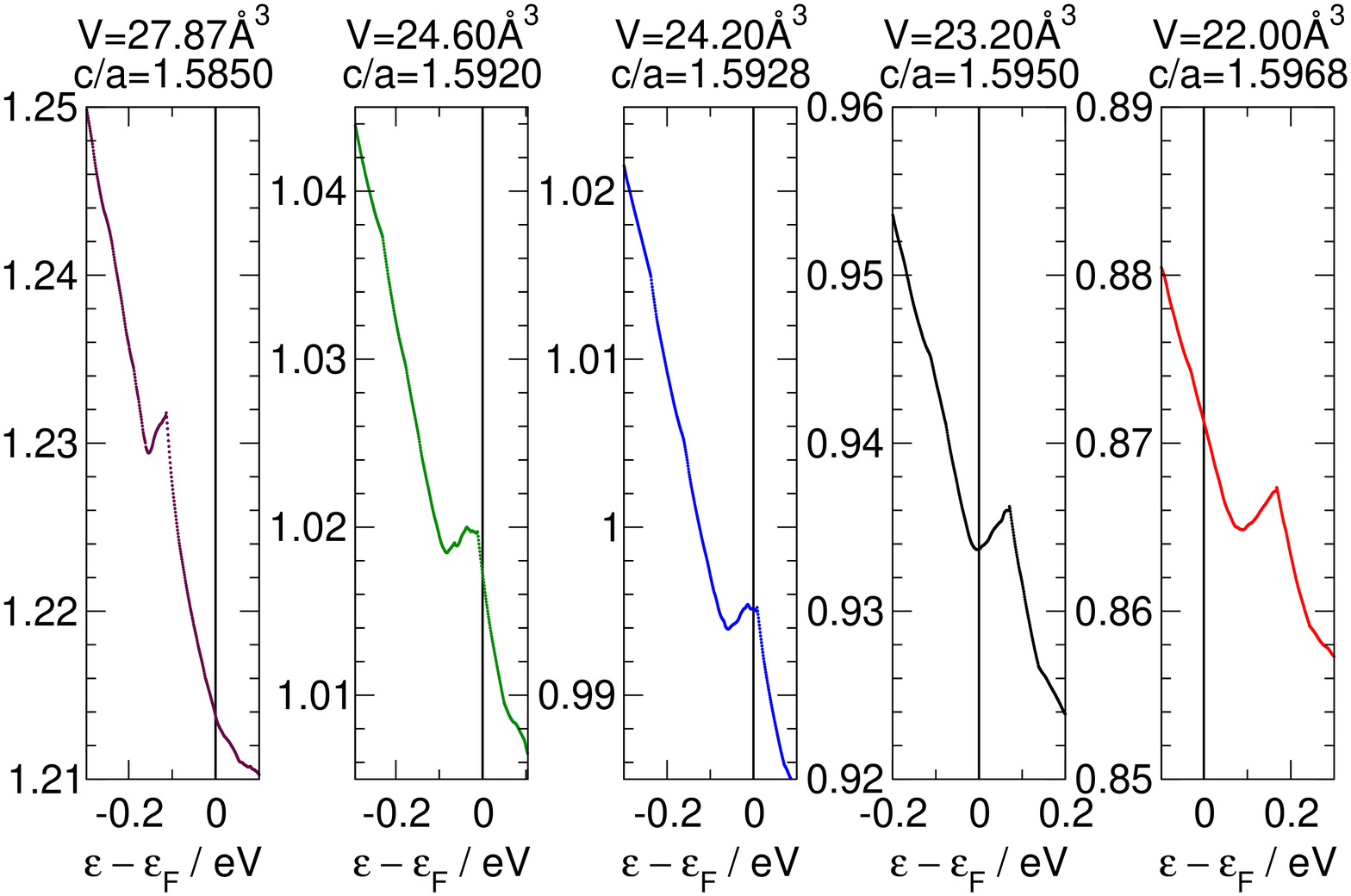 scaled 270}
\caption{(color online). Upper panel: Total density of states of Osmium at two 
different volumes. Maroon (black) full lines: zero pressure; 
	red (light-gray) dashed
lines: pressure of about 180 GPa. The Fermi level is situated in a
smooth valley of the DOS.
        Five lower panels, from left to right: magnification
	of the DOS around the Fermi level at zero pressure; at that
	pressure, where the ellipsoid at the $\Gamma$-point appears, at that
	pressure, where the neck is created, at that pressure, 
        where the ellipsoid
	at the L-point appears, and at a pressure of about 180 GPa.
        }\label{OsDOS}
\end{figure}

ETT should manifest themselves in van Hove singularities of the 
DOS passing through the Fermi level under pressure. 
The upper panel of
Figure \ref{OsDOS} presents the DOS of Osmium
at zero pressure and at about 180 GPa. The 5d band is considerably
broadened at such a high pressure, in comparison with the zero pressure
case. However, the Fermi level is situated
in a smooth valley of the DOS in the whole pressure range. 
In particular, no Fermi level crossing of any singularity is obvious
in this overview panel.

The lower part of Figure \ref{OsDOS} shows the DOS close to the
Fermi level in about hundredfold magnification, for five different
pressures including the three transition pressures and the two
limiting cases of the upper panel.
A tiny anomaly, caused by the Lifshitz transitions,
is found moving through the Fermi level.
Despite the huge number
of $96^3$ ${\bf k}$-points in the full BZ used for these calculations,
the estimated DOS resolution that can be achieved for the present
band dispersions by means of the linear
tetrahedron method amounts to 50 meV which is about the width
of the observed anomaly. Thus, no details of this feature can be resolved,
it remains unclear to which extent the individual ETT contribute to it.
(Note, that the shape is changing with pressure. This fact also
hints to resolution problems.)

\begin{figure}
\BoxedEPSF{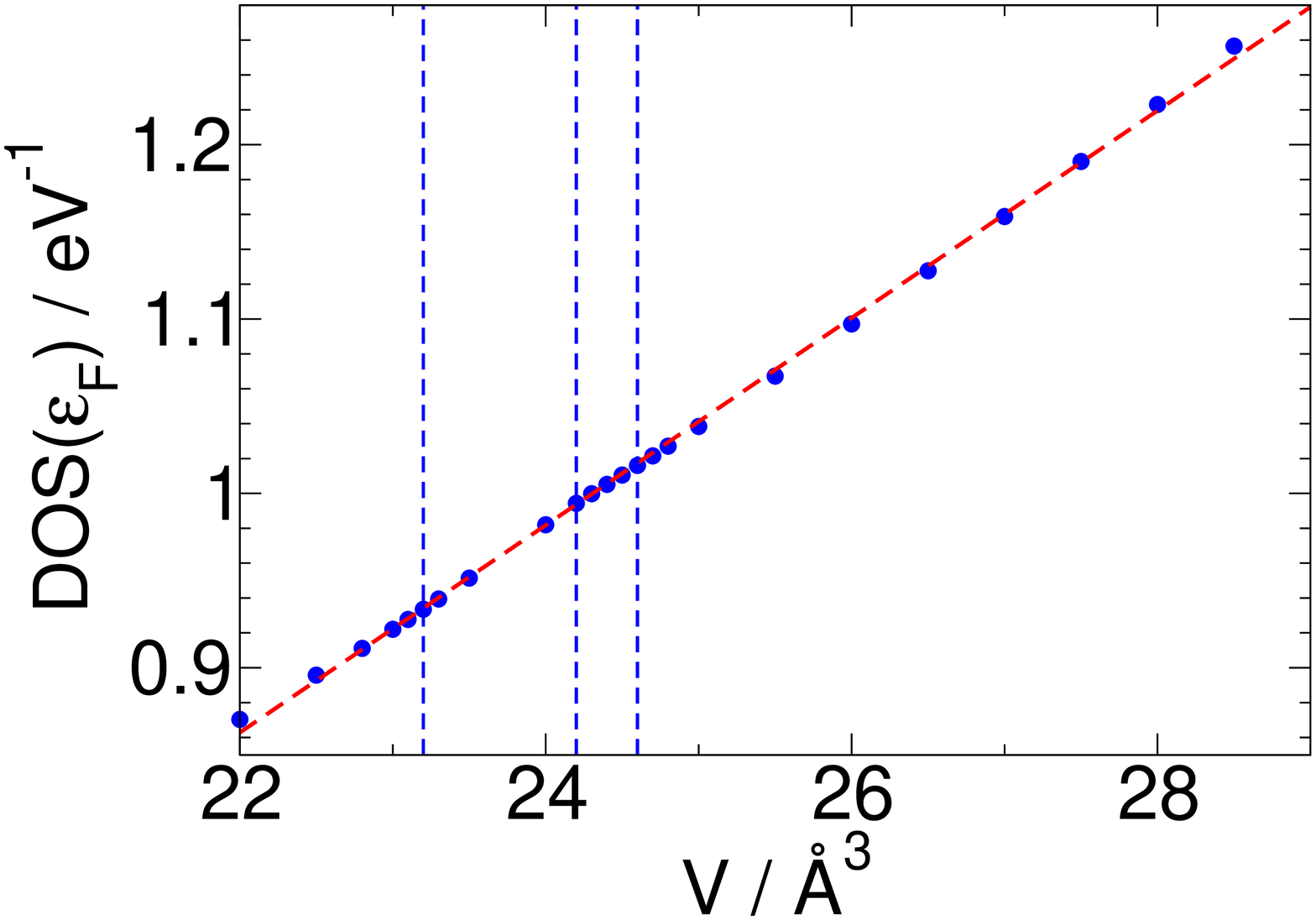 scaled 225}
\caption{
	(color online).
        DOS at the Fermi energy versus volume. 
        The red (gray) solid line shows a linear function
        fitted to the data. The blue (black) dashed lines mark those volumes, 
	where the Lifshitz transitions take place.
        }\label{presDOS}
\end{figure}

As the ETT anomaly amounts to about $10^{-3}$ 
of the total DOS, it is
probably impossible to detect it in the elastic properties. This
becomes even more clear if we consider the volume dependence of the
DOS at the Fermi energy, Figure \ref{presDOS},
which decreases almost linearly with unit
cell volume.
The red (gray) solid line shows a linear fit
to the data. No peculiarity is visible at the 
Lifshitz transitions, marked by blue (black) dashed lines. 
To understand this point, we have estimated the strength of the
van Hove singularity related to the
first ETT, the appearance of the ellipsoid at $\Gamma$.
The extra contribution to the DOS due to this ellipsoid can be calculated as
\begin{equation}
|\delta g(\varepsilon)| = \frac{V}{2\pi^2}
\sqrt{\frac{8m_xm_ym_z}{\hbar^6}}\sqrt{(\varepsilon_{cr}-\varepsilon)}
\end{equation} 
with $V=\unit[27.87]{\AA^3}$. Here, $\hbar^2/2m_i$ denotes the curvature of the
ellipsoid in direction $i$, which has been fitted to 
$\hbar^2/2m_x=\unit[171.4]{eV\AA^2}$,
$\hbar^2/2m_y=\unit[161.4]{eV\AA^2}$ and 
$\hbar^2/2m_z=\unit[97.1]{eV\AA^2}$. Then 
$\delta g(\varepsilon) \approx \unit[0.0009]{eV^{-1}}
\sqrt{(\varepsilon_{cr}-\varepsilon)/{\rm eV}}$ or about 
$\unit[0.00009]{eV^{-1}}$
at 10 meV distance from the position of the van Hove singularity, 
$\varepsilon_{cr}$. Here, the
crucial point is the tiny prefactor of the square-root singularity.
This extra contribution to the pressure
of the Fermi gas will not be visible in elastic data obtained
by today's experimental or numerical means.

Summarizing this section, we find that the disclosed Lifshitz
transitions are very weak and probably not detectable by
measurements of elastic properties. However, there might be a chance of 
observing anomalies in magneto-transport or thermopower under pressure.

\section{$c/a$ ratio under pressure}\label{sec_geom}

We now proceed to a comparison of several measured and calculated
variants of the pressure and volume
dependence of the $c/a$ ratio. A critical evaluation and data analysis
will help to clarify the question, if there is any anomaly in the
data close to 25 GPa (or 10 GPa), as proposed by several 
authors.\cite{Occelli04_095502,Ma05_174103,Sahu05_113106}

\begin{figure}
\BoxedEPSF{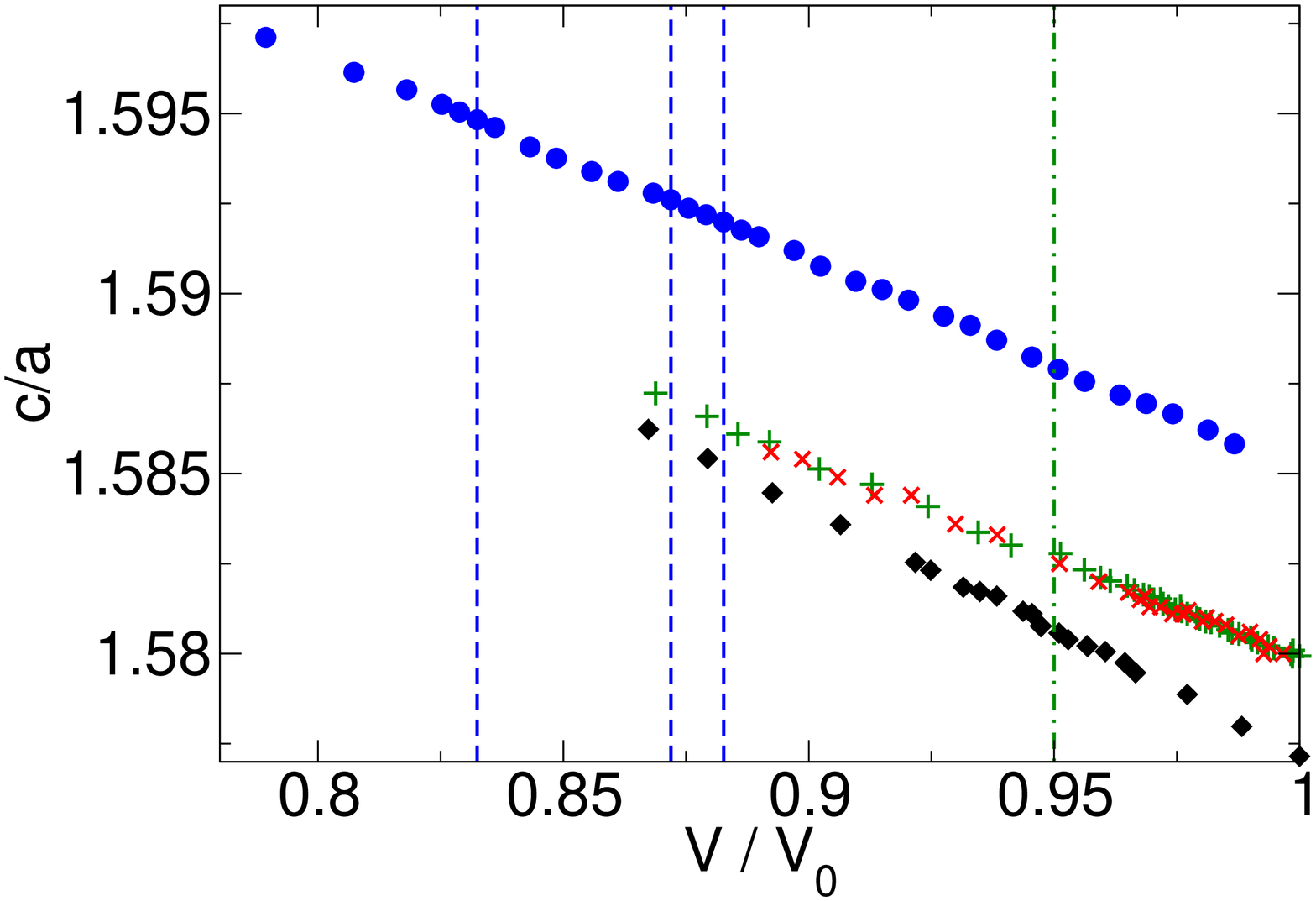 scaled 240}
\BoxedEPSF{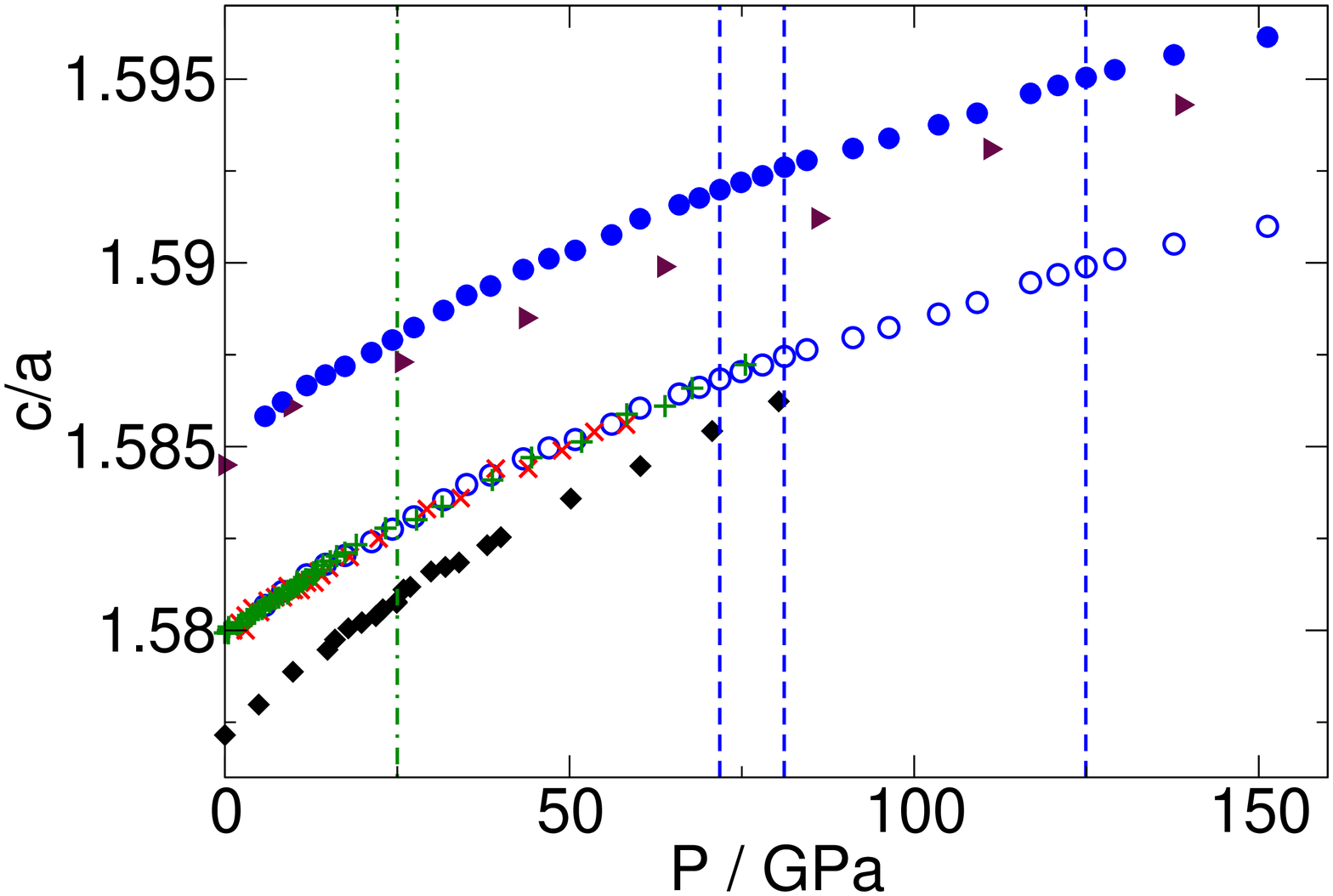 scaled 240}
\caption{(color online). Upper panel: relaxed $c/a$-ratio 
	against normalized volume;
	lower panel: relaxed $c/a$-ratio against pressure.
        The meaning of the symbols $+$, $\times$, $\bullet$ and green (gray)
	dashed-dotted line is the same as in Figure \ref{OsEPV},
	lower panel. Moreover,
	black diamonds denote calculations by Ma {\em et al}..
	\cite{Ma05_174103}
        Additional data from calculations by Sahu and Kleinman are 
        denoted by violet (black) triangles in the 
	lower panel.\cite{Sahu05_113106} The blue (black) open circles are 
	a shift of our data down by 0.00515. 
	Error-bars are below symbol size.
	The position of the Lifshitz transitions (blue (black) lines)
	in the lower graph was obtained from the FPLO EOS.}\label{cdaVP}
\end{figure}

Figure \ref{cdaVP} shows the dependence of $c/a$ on volume and
on pressure. Similar to the EOS case, both experimental data sets coincide
within the error estimates given in Refs.\ \onlinecite{Occelli04_095502}
and \onlinecite{Kenichi04_012101}. 
The three computed data sets run more or less
parallel with the experimental ones. Unlike the EOS data,
the offset of the LDA results  of about $0.005$ 
(blue (black) circles, present calculation,
and violet (black) triangles, calculation by Sahu and Kleinman)
is larger than the offset of the GGA results by Ma {\em et al.} (black
diamonds). Additionally, we have introduced another data set (blue (black)
open circles) in the lower panel of Figure \ref{cdaVP}, denoting our 
calculated data shifted down by 0.00515. These offset-corrected LDA data
show a remarkable coincidence with the measured data both in slope and 
curvature. Thus, they can be used to extrapolate the experimental data
into the high-pressure region in order to estimate experimental 
transition pressures given in Section \ref{FermiS}.

In $(c/a)(V/V_0)$ no anomaly is visible at any of the 
Lifshitz transition volumes
(blue (black) dashed lines), and no anomaly is visible 
either at a volume of about
$0.95 V_0$, corresponding to 25 GPa.
The proposed anomalies in $(c/a)(P)$ will be analyzed subsequently.
Before, we check the dependence of $c/a$ on the volume in detail, which looks
linear at first glance.

We performed linear fits to the individual data sets in the volume range
$0.87 < V/V_0 < 1$, accounting for the known uncertainties of the experimental 
values, and considering the theoretical data points with equal weight. 
To characterize the quality of the fits, Table \ref{Osstd} displays 
$\chi_n=\sqrt{\chi^2/(n_p-n_e)}$ where $\chi^2$ denotes the sum of the error
weighted square deviations, $n_p$ the number of the data points, and $n_e$
the number of adjustable parameters, here $2$. Moreover, for all data sets,
Table \ref{Osstd} gives estimates of the standard deviation $\sigma$
presuming validity of the linear dependence 
and uniform uncertainties of all data points included.
Two conclusions can be drawn from this check: (i) the dependence
of $c/a$ on volume is linear within experimental accuracy according to
the $\chi_n$ values,
and (ii)
the random errors of the theoretical data are comparable with the
experimental inaccuracy.
A linear fit of our calculation results for the whole volume range,
$0.78<V/V_0<1$, yields a $\sigma$ value of $8.7\cdot 10^{-5}$,
resulting from a slight non-linearity at higher compressions.

\begin{table}
\begin{tabular}{l|c|c|c}\hline \hline
    & $\chi_n$ & $\sigma$ & $V_0/$\AA$^3$\\ \hline
experiment (Occelli {\em et al.})\cite{Occelli04_095502}& $0.73$ & $7.4\cdot 10^{-5}$ & $27.949$\\ 
experiment (Takemura)\cite{Kenichi04_012101} & $1.22$ & $1.2\cdot 10^{-4}$ & $27.977$\\ \hline 
present calculation  & & $5.9\cdot 10^{-5}$ & $27.87$\\ 
calculation by Ma {\em et al.}\cite{Ma05_174103} & & $8.1\cdot 10^{-5}$ & $28.48$\\ 
\hline \hline
\end{tabular}
\caption{Quality of linear fits to the $(c/a)(V/V_0)$ data 
	displayed in Figure \ref{cdaVP}, upper panel, in the range
        $0.87<V/V_0<1$. For the definition of $\chi_n$ and $\sigma$ see text. 
	The fourth column displays the values used for $V_0$.
	Values of adjustable parameters are given in the Appendix.}
\label{Osstd}
\end{table}

Now we come to the key question of a possible anomaly in $(c/a)(P)$
at about 25 GPa, green (gray) dashed-dotted line in 
the lower panel of Figure \ref{cdaVP}.
We stress again, in accordance with Refs.\
\onlinecite{Occelli04_095502} and \onlinecite{Ma05_174103},
that there is no doubt about the linearity of $(c/a)(V/V_0)$
within experimental precision. Thus, if $(c/a)(P)$ would be
piecewise linear, as suggested by Occelli {\em et al.}, Sahu and Kleinman,
as well as Ma {\em et al.},
then $P(V/V_0)$ had to be piecewise linear as well.
In turn, this would imply a zero pressure derivative of the bulk modulus,
$B'=0$, in each of the pressure regions below and above the anomaly -
a clear contradiction with the EOS fits by Occelli {\em et al.} who
find $B'\approx 4$ in both regions.\cite{Occelli04_095502}

Given a non-zero pressure derivative of the bulk modulus,
$P(V)$ could most simply be approximated by a second-order polynomial.
Since $c/a$ is linear in the volume,
this approximation translates into a square-root dependence of 
$c/a$ on pressure:
\begin{equation}
  (c/a)(P) = (c/a)_m + d \sqrt{(P + P_m)} \; ,
\label{goodfit}
\end{equation}
where $(c/a)_m$, $d$, and $P_m$ are fit parameters.

It is clear, that a square-root behavior can be fitted by two
linear pieces within a restricted range.
Table \ref{OsPstd} gives a comparison of square-root and piece-wise
linear fits between zero and 60 GPa, which is the range covered by all of the
data sets. (We do not take into account the data
by Sahu and Kleinman, since they have only four points in the considered
pressure range.) According to Refs.\ \onlinecite{Occelli04_095502} and
\onlinecite{Ma05_174103}
we assumed the kink between the linear pieces at the suggested 25 GPa and 
27 GPa, respectively.
 
Table \ref{OsPstd} shows that both kinds of fits work 
almost equally well, with a small advantage
of the square-root fit.
Additionally for the data by Occelli {\em et al}., we performed a four 
parameter piece-wise linear fit, allowing the critical pressure $P_c$ to vary:
Only a slightly better approximation was obtained ($\chi_n=0.51$ for the
complete data set), but $P_c$ amounts to $19\pm2$ GPa. Finally
we have checked parabolic and cubic fits, which
yield very similar $\chi_n$ and $\sigma$ values 
compared to the square-root fit.

According to Tab.\ \ref{OsPstd} and the non-vanishing pressure derivative
of the compressibility
argument above, we believe that there is not enough evidence to 
conclude an anomaly at about 25 GPa.
Regarding the DFT calculations,
no peculiarity in the band structure or Fermi surface topology is
observed in this pressure range. Thus, it would be artificial to
expect an anomaly in the computed elastic data. Of course,
regarding the experiments, other than electronic origins of an anomaly
cannot be excluded. However, in our opinion, such an anomaly is not proved by
in the available data sets. Error correlations mentioned
in Ref.\ \onlinecite{Occelli04_095502} could be caused by the change of the 
experimental setup at $\unit[26]{GPa}$.
\begin{table}
\begin{tabular}{l|c|c|c|c|c|c}\hline \hline
		& \multicolumn{2}{c|}{Eq.\ (\ref{goodfit})} & \multicolumn{4}{c}{piece-wise linear function} \\ \hline
    &    & 	& \multicolumn{2}{c|}{kink: $\unit[25]{GPa}$} & \multicolumn{2}{c}{kink: $\unit[27]{GPa}$} \\ \hline
  & $\chi_n$ & $\sigma$ & $\chi_n$ & $\sigma$ & $\chi_n$ & $\sigma$ \\ \hline
exp. (Occelli {\em et al.})\cite{Occelli04_095502} & $0.69$ & $6.8\cdot 10^{-5}$ & $0.71$ & $7.4\cdot 10^{-5}$  & $0.85$ & $8.5\cdot 10^{-5}$\\
exp. (Takemura)\cite{Kenichi04_012101} & $1.27$ & $1.3\cdot 10^{-4}$ & $1.28$ & $1.3\cdot 10^{-4}$ & $1.26$ & $1.3\cdot 10^{-4}$\\ \hline
present calculation		& & $7.5\cdot 10^{-5}$ & & $1.1\cdot 10^{-4}$ & & $1.0\cdot 10^{-4}$ \\ 
calc. by Ma {\em et al.} \cite{Ma05_174103}	& & $7.8\cdot 10^{-5}$ & & $1.0\cdot 10^{-4}$ & & $9.7\cdot 10^{-5}$\\
\hline \hline
\end{tabular}
\caption{Quality of different fits to the data displayed in
	Figure \ref{cdaVP}, lower panel, in the pressure range between zero
        and 60 GPa. For the definition of $\chi_n$ and $\sigma$ see text.
        Values of adjustable parameters are given in the Appendix.}
\label{OsPstd}
\end{table}
\section{Summary}
We predict
three Lifshitz transitions to occur in hexagonal
close-packed Osmium under pressure up to 180 GPa.
The corresponding van Hove singularities in the density 
of states are probably too small to produce any measurable effect
in the elastic properties but may be detectable in transport properties. 
Since the lowest critical pressure amounts to $70 \ldots 80$ GPa,
it is not related with a previously suggested kink at 25 GPa in the pressure
dependence of $c/a$.
We demonstrate that this kink can
be replaced by a smooth dependence
without any change in the statistic
significance. 
Thus, we cannot confirm the electronic origin of the proposed anomaly. 
\section*{Acknowledgments}

We are indebted to Ulrike Nitzsche, Ingo Opahle, Norbert Mattern, and
Ulrich Schwarz for enlightening discussions.
Yanming Ma, Daniel L. Farber, B.R. Sahu, and Leonard Kleinman kindly
provided numerical data from their related publications. Financial support by
DFG, SPP 1145, is gratefully acknowledged.
\appendix
\section*{Appendix}
\subsection{Linear fit to $(c/a)(V/V_0)$}
\centerline{$c/a = A\cdot (V/V_0) + B$}
\vspace{0.1cm}
\begin{tabular}{l|c|c}\hline \hline
						   & $A$	 & $B$ \\ \hline
exp. (Occelli {\em et al.})\cite{Occelli04_095502} & $-0.0539(4)$ & $1.6339(4)$ \\
exp. (Takemura)\cite{Kenichi04_012101}		   & $-0.0540(6)$ & $1.6339(6)$ \\ \hline
present calculation				   & $-0.0594(2)$ & $1.6445(1)$ \\
calc. by Ma {\em et al.} \cite{Ma05_174103}	   & $-0.0678(6)$ & $1.6450(6)$ \\ \hline \hline
\end{tabular}

\subsection{Square-root fit to $(c/a)(P)$}
\begin{tabular}{l|c|c|c}\hline \hline
						   & $(c/a)_m$	& $d/$GPa$^{-1/2}$	      & $P_m/$GPa \\ \hline
exp. (Occelli{\em et al.})\cite{Occelli04_095502} & $1.572(1)$  & $0.00146(9)$ & $28(5)$ \\
exp. (Takemura)\cite{Kenichi04_012101}		   & $1.572(2)$  & $0.0015(1)$  & $30(9)$ \\ \hline
present calculation				   & $1.5753(8)$ & $0.00162(5)$ & $37(4)$ \\
calc. by Ma {\em et al.} \cite{Ma05_174103}	   & $1.5708(10)$ & $0.00157(8)$ & $16(4)$ \\ \hline \hline
\end{tabular}
\subsection{Piece-wise linear fit to $(c/a)(P)$}
\subsubsection{Kink at $\unit[25]{GPa}$}
\noindent
A - $c/a$ at $P=\unit[0]{GPa}$\\
B - slope between $\unit[0]{GPa}$ and $\unit[25]{GPa}$\\
C - slope between $\unit[25]{GPa}$ and $\unit[60]{GPa}$\\~\\
\begin{tabular}{l|c|c|c}\hline \hline
						   & $A$	& $10^4B/$GPa$^{-1}$	     & $10^4C/$GPa$^{-1}$ \\ \hline
exp. (Occelli {\em et al.})\cite{Occelli04_095502} & $1.579933(9)$ & $1.23(2)$ & $0.80(3)$ \\
exp. (Takemura)\cite{Kenichi04_012101}		   & $1.57995(4)$ & $1.17(3)$  & $0.86(3)$ \\ \hline
present calculation				   & $1.58525(2)$ & $1.117(8)$ & $0.955(6)$\\
calc. by Ma {\em et al.} \cite{Ma05_174103}	   & $1.57729(7)$ & $1.46(3)$  & $1.06(4)$ \\ \hline \hline
\end{tabular}
\subsubsection{Kink at $\unit[27]{GPa}$}
\noindent
A - $c/a$ at $P=\unit[0]{GPa}$\\
B - slope between $\unit[0]{GPa}$ and $\unit[27]{GPa}$\\
C - slope between $\unit[27]{GPa}$ and $\unit[60]{GPa}$\\~\\
\begin{tabular}{l|c|c|c}\hline \hline
						   & $A$	& $10^4B/$GPa$^{-1}$	     & $10^4C/$GPa$^{-1}$ \\ \hline
exp. (Occelli{\em et al.})\cite{Occelli04_095502} & $1.579935(2)$ & $1.219(14)$ & $0.78(4)$ \\
exp. (Takemura)\cite{Kenichi04_012101}		   & $1.57995(4)$ & $1.16(3)$ & $0.84(3)$ \\ \hline
present calculation				   & $1.58525(2)$ & $1.113(8)$ & $0.940(7)$\\
calc. by Ma {\em et al.} \cite{Ma05_174103}	   & $1.57730(6)$ & $1.45(3)$ & $1.02(4)$ \\ \hline \hline
\end{tabular}\\
\vspace{4cm}\\

\end{document}